\documentclass[preprint,12pt, nonatbib]{elsarticle}

\makeatletter
\def\ps@pprintTitle{%
   \let\@oddhead\@empty
   \let\@evenhead\@empty
   \let\@oddfoot\@empty
   \let\@evenfoot\@oddfoot
   \let\c@author\@empty
}
\let\c@author\relax  
\let\c@bibliography\relax  
\makeatother

\usepackage[backend=biber,style=nature]{biblatex}
\usepackage{booktabs}
\usepackage{amssymb}
\usepackage{xcolor}
\usepackage{amsthm}
\usepackage{mathtools}
\usepackage{enumitem}
\usepackage{graphicx}
\usepackage{subfigure}
\usepackage{autobreak}
\usepackage{lineno}

\usepackage[margin=2cm]{geometry}
\usepackage{multirow}
\usepackage{floatrow}
\usepackage{colortbl}
\usepackage{threeparttable}
\usepackage{booktabs}
\usepackage{pdflscape}

\usepackage{booktabs,threeparttable,array}
\newcolumntype{Y}{>{\centering\arraybackslash}p{1.15cm}}

\usepackage{xurl}
\usepackage[colorlinks=true,linkcolor=black,citecolor=black,urlcolor=blue]{hyperref}

\usepackage{graphicx}
\usepackage{subcaption}
\usepackage{comment}
\usepackage{multirow}
\newfloatcommand{capbtabbox}{table}[][\FBwidth]

\definecolor{orange}{rgb}{1,0.5,0}
\definecolor{red}{RGB}{198,0,35}
\definecolor{amberseldef}{rgb}{1.0, 0.49, 0.0}
\definecolor{ceruleanblue}{rgb}{0.16, 0.32, 0.75}
\definecolor{amber}{rgb}{1.0, 0.49, 0.0}
\definecolor{dodgerblue}{rgb}{0.12, 0.56, 1.0}
\definecolor{pureblue}{rgb}{0, 0, 1.0}

\definecolor{blue}{rgb}{0.0, 0.28, 0.67}

\def\hmath$#1${\texorpdfstring{{\rmfamily\textit{#1}}}{#1}}

\AtBeginDocument{\hypersetup{citecolor= pureblue,linkcolor = pureblue,urlcolor = dodgerblue}}

\begin{document}


\begin{frontmatter}

\title{Unveiling Uniform Shifted Power Law in Stochastic Human and Autonomous Driving Behavior}

\address[1]{Department of Civil and Environmental Engineering, University of Wisconsin-Madison, Madison, Wisconsin, 53706, USA}

\cortext[cor1]{*Corresponding author: xli2485@wisc.edu}
\fntext[fn1]{\dag~Wang Chen and Heye Huang contributed equally to this work.}

\author[1]{Wang Chen\textsuperscript{\dag}}
\author[1]{Heye Huang\textsuperscript{\dag}}
\author[1]{Ke Ma}
\author[1]{Hangyu Li}
\author[1]{Shixiao Liang}
\author[1]{Hang Zhou}
\author[1]{Xiaopeng Li\corref{cor1}}

\begin{abstract}
Accurately simulating rare but safety-critical driving behaviors is essential for the evaluation and certification of autonomous vehicles (AVs). However, current models often fail to reproduce realistic collision rates when calibrated on real-world data, largely due to inadequate representation of long-tailed behavioral distributions. Here, we uncover a simple yet unifying shifted power law that robustly characterizes the stochasticity of both human-driven vehicle (HV) and AV behaviors, especially in the long-tail regime. The model adopts a parsimonious analytical form with only one or two parameters, enabling efficient calibration even under data sparsity. Analyzing large-scale, micro-level trajectory data from global HV and AV datasets, the shifted power law achieves an average $\mathrm{R}^2 \approx 0.97$ and a nearly identical tail distribution, uniformly fits both frequent behaviors and rare safety-critical deviations, significantly outperforming existing Gaussian–based baselines. When integrated into an agent–based traffic simulator, it enables forward-rolling simulations that reproduce realistic crash patterns for both HVs and AVs, achieving rates consistent with real-world statistics and improving the fidelity of safety assessment without post hoc correction. This discovery offers a unified and data–efficient foundation for modeling high-risk behavior and improves the fidelity of simulation-based safety assessments for mixed AV/HV traffic. The shifted power law provides a promising path toward simulation-driven validation and global certification of AV technologies.
\end{abstract}

\end{frontmatter}

\section{Introduction}\label{sec: 1}

Safety is a critical priority in the development and deployment of high-level autonomous vehicles (AVs) \cite{NHTSA2024,kalra2016driving}. Many recent studies have attempted to evaluate AV safety by simulating interactions between the ego AV and surrounding human-driven vehicles (HVs), using human driver behavior models (and corresponding vehicle dynamics) such as car-following, lane-changing, and interactions with traffic control devices \cite{derbel2013modified,ahmed1996models,kesting2007general}. Simulation serves as a faster, more controlled, repeatable, and cost-effective alternative to real-world testing, allowing for comprehensive evaluation and comparison of autonomous driving systems \cite{paden2016survey,bansal2017forecasting,koopman2017autonomous}.
However, to the authors' knowledge, few traffic simulators can yield realistic collision rates of the correct order of magnitude, directly from driver behavior models calibrated using real-world data \cite{kaur2021survey,sun2020scalability}. Recently, a plausible effort was made to use traffic simulation to predict reasonable collision rates~\cite{yan2023learning}; however, those rates were obtained only after applying an acceptance rate to the model’s direct collision predictions, spanning several orders of magnitude. This leads to an intriguing research question: why can’t behavior models calibrated with real-world road traffic data accurately reproduce collision risks?

We notice that the collision counts of most non-collision-free traffic simulators are essentially determined by the stochasticity of their underlying driver behavior models. As shown in Figure~\ref{fig:fig1}(d), the classic modeling approach represents behavioral stochasticity through a predefined uncertainty distribution and then calibrates its parameters using real-world trajectory data~\cite{o2018scalable, riedmaier2020survey, hakobyan2020learning, ellis2009modelling, uugurel2024correcting}. 
Typical formulations include Gaussian and its variant models that assume symmetric deviations from nominal trajectories~\cite{hakobyan2020learning, kotz2012laplace, ahsanullah2014normal}, uniform or bounded noise models that constrain perturbations within fixed limits~\cite{riedmaier2020survey}, and logistic or Boltzmann-rational (softmax-based) choice models that describe decision uncertainty through exponential-family distributions~\cite{fridovich2020confidence}. The calibration objective in these models is typically to minimize the discrepancy between simulated and observed trajectories, such as speed, spacing, or acceleration profiles. While such models can reproduce traffic characteristics well in common safe-driving scenarios, their performance in risky scenarios is limited. This limitation arises because observed naturalistic data are abundant in low-risk conditions but become increasingly scarce in high-risk, long-tail scenarios. As a result, the calibrated models tend to fit high-frequency normal behaviors yet drastically underestimate the probability of rare but safety-critical scenarios \cite{al2024review,fildes2015effectiveness}. Consequently, traffic simulators using these models often fail to reproduce realistic collision rates, leaving a significant modeling gap in the tail distribution of vehicle behaviors.

In other words, existing driver behavior models likely overlook the long-tail characteristics of risky scenarios that are critical for safety assessment. For example, as revealed in \cite{cheng2021limits}, the normal distribution structure, adopted in many relevant studies \cite{wiest2012probabilistic,ziegler2022modeling}, when calibrated with available data, may significantly underestimate conflicts in risky scenarios. This could be a possible reason why \cite{yan2023learning} needed to revise collision rates to offset this underestimation. However, this fundamental issue has not yet received much attention in the research community. It is still unknown which structure can best capture the tail distribution of realistic driving behavior when calibrated only with available data that are highly skewed toward the normal behavior range.

This study aims to investigate this fundamental challenge: Is there a parsimonious distribution pattern that can produce a reasonable driving behavior tail distribution with available data? To our surprise, we discover a simple shifted power law model with one- or two-parameter form that very well predicts the tail distribution of risky driving behavior. The proposed model fits uniformly well several HV and AV datasets in different driving environments across several continents, including highD (Germany, HV)~\cite{krajewski2018highd}, CitySim (USA, HV)~\cite{zheng2024citysim}, and a comprehensive collection of multiple AV datasets, including Waymo~\cite{sun2020scalability,hu2022processing}, Argoverse~2~\cite{wilson2023argoverse}, MicroSimACC~\cite{yang2024microsimacc}, CATS~\cite{shi2021empirical,ma2025real}, OpenACC~\cite{makridis2021openacc}, and Central Ohio ACC~\cite{xia2023automated}, spanning highway, urban, and mixed driving conditions. We show that the proposed model achieves accurate fits not only for common driving patterns but also for rare safety-critical tail events. Furthermore, simulation-based crash rate analysis shows that the proposed shifted power law yields results consistent with empirical crash statistics, measured per million vehicle-miles traveled (VMT), narrowing the gap between simulated and naturalistic driving environments.  This discovery advances our knowledge of tail driving behavior distributions of AV/HV, critical to safety performance. It opens a gateway to enable simulation directly calibrated with real-world data to accurately predict risky behavior, tail distribution, and probabilistic collision risks of mixed AV/HV traffic in different countries across the world. The model could be used to formulate solutions to outstanding challenges of AV assessment, including the ``curse of rarity'' (\emph{e.g.}, via accurate tail distribution prediction) and the ``curse of dimensionality'' (\emph{e.g.}, via the parsimonious analytical model structure). This model can be further extended to different AV manufacturers and diverse driving scenarios, and thus paves the way for a comprehensive quantitative benchmark for AV risk assessment and certification.


\section{Results}\label{sec: 2}

\subsection{Modeling stochastic behavior with shifted power law}\label{sec: 2.1}

\begin{figure}[!ht]
\centering
\includegraphics[width=1\linewidth]{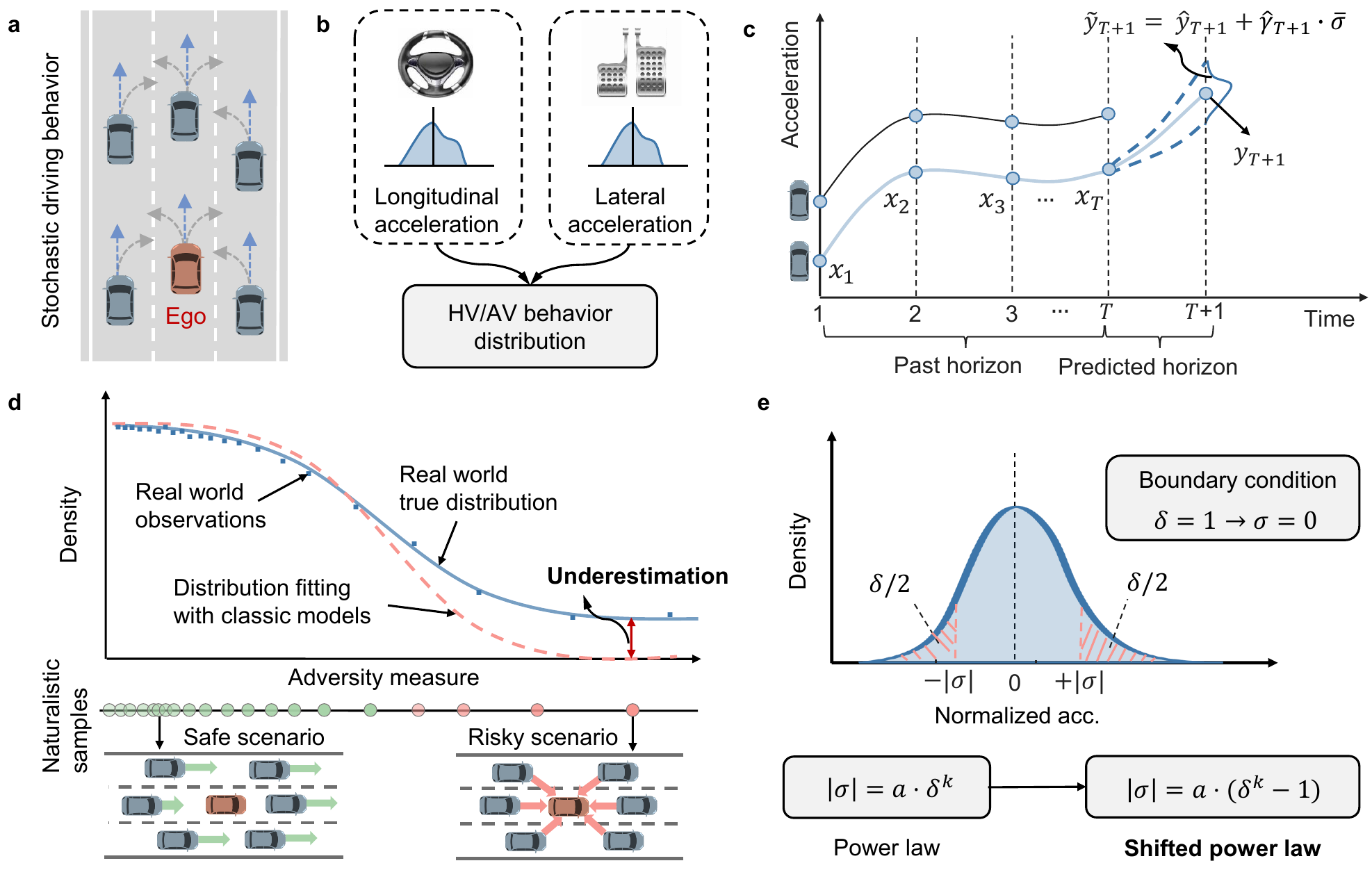}
\caption{\textbf{Modeling stochastic driving behavior with the shifted power law.}
\textbf{a}, Illustration of stochastic driving behavior: the ego vehicle interacts with surrounding vehicles through both lateral and longitudinal directions.
\textbf{b}, Decoupling the longitudinal and lateral acceleration to model stochastic driving behavior of an HV/AV.
\textbf{c}, The AI model predicts the mean acceleration $\hat{y}_{T+1}$ and standard deviation $\hat{\gamma}_{T+1}$ based on the previous states $x_{1:T}$. Then we focus on the normalized residual distribution $\bar{\sigma} := (y_{T+1} - \hat{y}_{T+1})/\hat{\gamma}_{T+1}$, where $y_{T+1}$ is the observation at time $T+1$. We assume $\bar{\sigma}$ is state-independent and thus the time index is dropped. 
\textbf{d}, Although classic models adequately approximate driving behavior in safe scenarios, which accounts for the majority of real-world observations, they systematically underestimate rare high-risk long-tail driving behaviors.
\textbf{e}, The shifted power law analytically links each possible threshold value $\sigma$ of the normalized residual random variable $\bar{\sigma}$, to the corresponding violation probability $\delta:=\mathbb{P}(|\bar{\sigma}|>|\sigma|)$ enabling interpretable risk modeling and realistic reproduction of safety-critical heavy tails. 
Together, Figures~\textbf{a–e} illustrate how the proposed model bridges microscopic behavioral stochasticity and macroscopic tail-risk characterization, providing a unified framework for quantitative analysis of HV/AV driving behaviors.}
\label{fig:fig1}
\end{figure}

We formulate the stochastic driving behavior model of a vehicle (either HV or AV) as a specific shifted power law. As illustrated in Figure~\ref{fig:fig1}(a), a vehicle running on a road continuously exhibits stochastic driving behavior. For the convenience of the analysis, as depicted in Figure \ref{fig:fig1}(b), we decouple the driving behavior into the longitudinal direction (governing acceleration and deceleration) and the lateral direction (governing lane-changing and lateral positioning).

Now we formulate the decoupled stochastic driving behavior. Consider a set of consecutive time points $t\in\{1, 2, \dots, T, T+1\}$, where $T$ is the current time step and $T+1$ is the next time step for prediction. At each time $t$, the state of the ego vehicle (an HV or AV) is denoted by $x_t$, which includes position, velocity, acceleration, spacing, lane ID, and relative positions with surrounding vehicles. As shown in Figure~\ref{fig:fig1}(c), given a sequence of past observations $x_{1:T}$, We assume that acceleration $\tilde{y}_{T+1}$ at prediction time step $T+1$ is stochastic and follows a distribution with mean  $\hat{y}_{T+1}$ and standard deviation $\hat{\gamma}_{T+1}$. We assume that $\hat{y}_{T+1}$ and $\hat{\gamma}_{T+1}$ are determined by past observations $x_{1:T}$ and can be learned with a neuro network model $g^{\mathrm{AI}}$ (see Section~\ref{sec: prediction} for details):
\begin{equation}
\hat{y}_{T+1}, \hat{\gamma}_{T+1} = g^{\mathrm{AI}}(x_{1:T}).
\label{eq:ai_pred}
\end{equation}
To simplify the analysis without loss of generality, we normalize the acceleration using the predicted mean and standard deviation:
\begin{equation}
\bar{\sigma}_{T+1} := \frac{y_{T+1} - \hat{y}_{T+1}}{\hat{\gamma}_{T+1}},
\label{eq:normalized_dev}
\end{equation}
where $y_{T+1}$ is the observed acceleration at time $T+1$, and $\bar{\sigma}_{T+1}$ is the normalized residual at time $T+1$. We assume that the normalized residual $\bar{\sigma}_{T+1}$ is state-independent since we have captured the state when predicting the mean and standard deviation. Consequently, we omit the subscript and use $\bar{\sigma}$ in the following text. This normalization allows us to approximate the original complex driving behavior distribution $\tilde{y}_{T+1}$ through a simple reconstruction:
\begin{equation}\label{eq: y_tilde}
    \tilde{y}_{T+1} = \hat{y}_{T+1} + \hat{\gamma}_{T+1}\bar{\sigma}.
\end{equation}

Classical stochastic driving models frequently employed Gaussian distributions or related variants to approximate the behavior of both HVs and AVs \cite{xu2014motion, wiest2012probabilistic, ziegler2022modeling, yan2023learning}. As shown in Figure \ref{fig:fig1}(d), these models can adequately characterize driving behavior in safe scenarios, where the vast majority of observations are concentrated near the mean, resulting in high-probability, low-risk events. The Gaussian assumption is tractable and often provides a reasonable fit for this central region of the distribution. However, this approach proves critically inadequate for modeling risky scenarios. In these cases, the true driving behavior distribution often exhibits a long-tail property, representing low-probability but high-consequence events, such as harsh braking or aggressive acceleration. Classical models with exponentially decaying tails, like the Gaussian distribution, drastically underestimate the probability of such extreme events \cite{cheng2021limits}. For example, if the normalized residual is assumed to follow a standard Gaussian distribution $\mathcal{N}(0,1)$, the probability of an event exceeding 5 standard deviations ($|\bar{\sigma}| > 5$) is on the order of $10^{-7}$. This infinitesimal probability can be several orders of magnitude lower than the true likelihood observed in real-world driving data, resulting in a significant and potentially dangerous underestimation of risk (see Section \ref{sec: 2.2} for details).

To accurately capture the long-tail driving behavior of an HV/AV, we propose a novel distribution model for the distribution $\mathcal{P}$ of $\bar{\sigma}$. As illustrated in Figure~\ref{fig:fig1}(e), we define a violation rate function $\delta(\sigma)$ as the probability of an observation of $\bar{\sigma}$ going beyond a given confidence interval $[-|\sigma|, +|\sigma|]$:
\begin{equation}
\delta (\sigma) :=\mathbb{P}(|\bar{\sigma}|>|\sigma|), \quad  \forall \sigma \in \mathbb{R}.
\label{eq:violation_rate}
\end{equation}
With this manipulation, we discover a shifted power law to describe the relationship between the confidence interval and the violation rate function $\delta(\sigma)$, as follows:
\begin{equation}
    |\sigma| = a(\delta^k-1), \quad \forall \sigma \in \mathbb{R},
    \label{eq:shifted_power_law}
\end{equation}
where $a > 0$ is the scale parameter and $k < 0$ is the decay exponent controlling tail heaviness. This formulation is inspired by the standard power law $|\sigma| = a\delta^k$, but is shifted to satisfy the logical boundary condition: when the confidence interval is zero ($\sigma = 0$), the violation rate must be one ($\delta = 1$).

Furthermore, based on the shifted power law, we can obtain the following cumulative distribution function (CDF) of normalized residual $\bar{\sigma}$:
\begin{equation}
F^{\mathcal{P}}(\sigma) =
\begin{cases}
\frac{1}{2} \left( 1 - \frac{\sigma}{a} \right)^{\frac{1}{k}}, & \sigma < 0;\\
1-\frac{1}{2} \left( 1 + \frac{\sigma}{a} \right)^{\frac{1}{k}}, & \sigma \geq 0.
\end{cases}
\label{eq:scaling_cdf}
\end{equation}
The corresponding probability density function (PDF) is:
\begin{equation}
f^{\mathcal{P}}(\sigma) = -\frac{1}{2ak} \left( 1 + \frac{|\sigma|}{a} \right)^{\frac{1}{k} - 1}, \quad \forall \sigma \in \mathbb{R}.
\label{eq:scaling_pdf}
\end{equation}
The detailed derivation of these analytical forms is presented in Section \ref{sec: derivation}, which provides the theoretical foundation for the following empirical validation.

The proposed shifted power law offers a concise and practical framework for modeling stochastic driving behavior. The model is characterized by only two parameters: a scale parameter $a$ and a decay exponent $k$. Furthermore, we discovered that the parameter $a$
can be fixed while still maintaining a high degree of accuracy, simplifying the model to a single-parameter representation in practice. This parsimony makes the shifted power law both computationally efficient and straightforward to apply in real-world autonomous driving systems. Unlike existing assumptions that may underestimate long-tail risks, the shifted power law enables a unified model of both frequent safe behaviors and rare risky behaviors, thereby improving the fidelity of safety assessment in mixed traffic involving both HVs and AVs. For the ease of parameter estimation, we apply a logarithmic transformation to Equation \eqref{eq:violation_rate}:
\begin{equation}
\log\left( 1 + \frac{|\sigma|}{a} \right) = k \log(\delta),
\label{eq:log_transform}
\end{equation}
which yields a linear relationship in the log-log space that enables efficient fitting using real-world data. This linear form enables robust and efficient fitting of the model to empirical data.

\subsection{Performance analysis of shifted power law}\label{sec: 2.2}

\begin{figure}[!ht]
\centering
\includegraphics[width=1\linewidth]{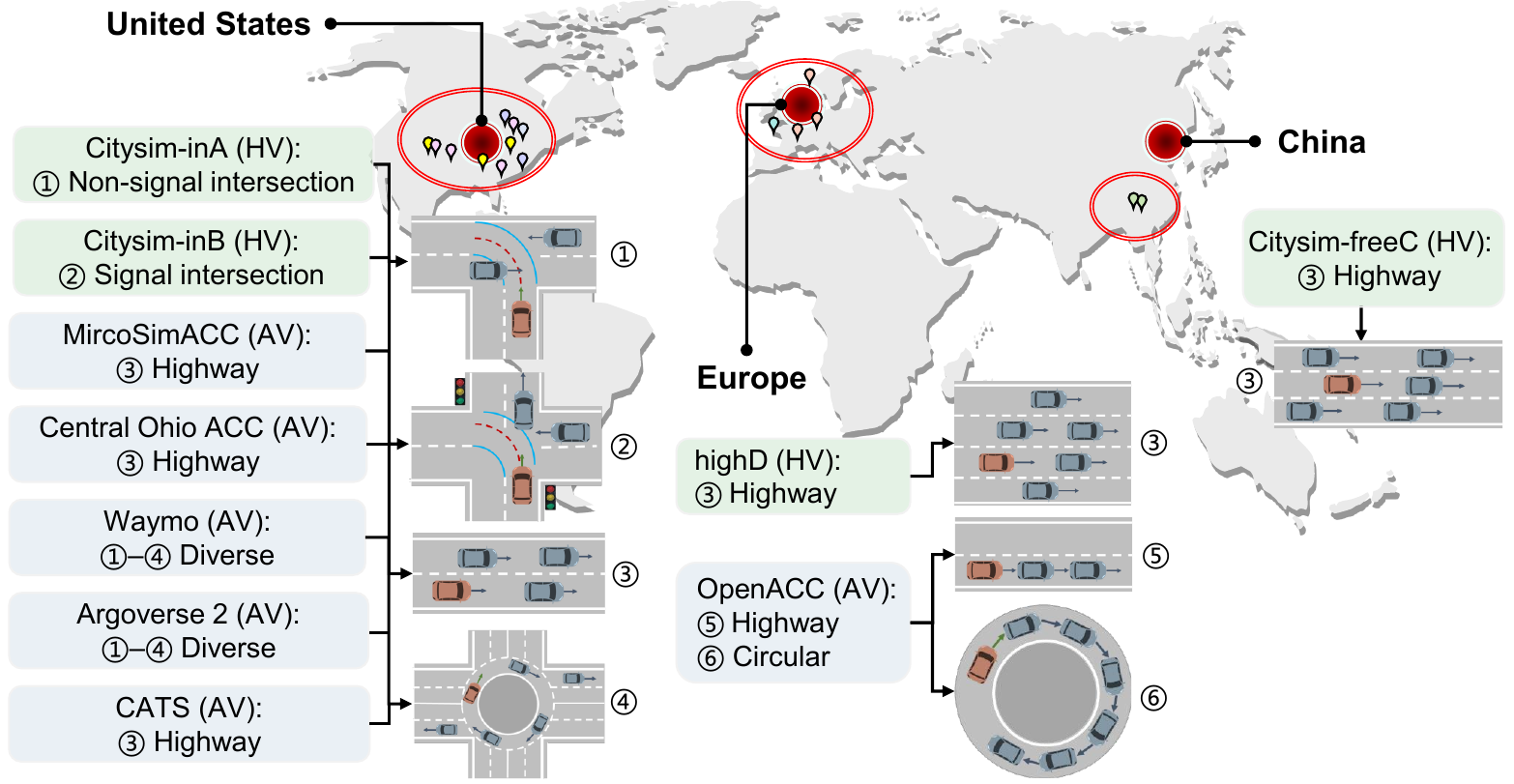}
\caption{\textbf{Validation of the shifted power law across global AV and HV datasets.} 
Datasets span three major regions: 
\textbf{United States}, including CitySim intersections (signalized and non-signalized) for HV, and multiple AV datasets such as MicroSimACC, Central Ohio ACC, Waymo, Argoverse~2, and CATS, covering highway and diverse driving environments. The third driving environment (i.e., highway) includes freeways and contains two, three, or four lanes in each direction.
\textbf{Europe}, including the highD dataset (HV highway) and the OpenACC dataset (AV highway). The OpenACC dataset records the car-following behavior of ego vehicles in the platoon on freeways (the fifth driving environment) and circular tracks (the sixth driving environment).
\textbf{China}, including the CitySim-freeC dataset (HV highway).}
\label{fig:fig2}
\end{figure}

We empirically validate the presence of heavy-tailed behavior in both HV and AV data. To ensure broad coverage and real-world fidelity, we incorporate trajectory data from multiple continents, covering both controlled AV experiments and large-scale naturalistic HV recordings. As summarized in Figure~\ref{fig:fig2}, the datasets span the United States, Europe, and China. In the United States, HV behavior is represented by the CitySim dataset at both signalized and non-signalized intersections~\cite{zheng2024citysim}, while AV behavior is captured across diverse platforms, including MicroSimACC~\cite{yang2024microsimacc}, CATS~\cite{shi2021empirical,ma2025real}, Central Ohio ACC~\cite{xia2023automated}, Waymo~\cite{sun2020scalability,hu2022processing}, and Argoverse~2~\cite{wilson2023argoverse}, spanning highway, urban, and mixed driving conditions. In Europe, HV trajectories are obtained from the highD dataset on German highways~\cite{krajewski2018highd}, complemented by the OpenACC dataset (with test sites in Italy, Sweden, and Hungary) for AV platoon highway and circular driving~\cite{makridis2021openacc}. In China, the CitySim-freeC dataset provides HV highway trajectories~\cite{zheng2024citysim}. Together, these datasets cover a broad spectrum of geographic regions, vehicle platforms, and driving environments, ensuring that the shifted power law is validated under diverse traffic conditions and interaction scenarios for both HV and AV driving behaviors. Details are depicted in Section \ref{sec: 4.1}.

\begin{table}[!ht]
\centering
\footnotesize
\caption{Shifted power law fitting results across AV and HV datasets.}
\label{tab:scaling_table}
\renewcommand{\arraystretch}{1.5}
\setlength{\tabcolsep}{3pt}

\begin{threeparttable}
\begin{tabular*}{\textwidth}{@{\extracolsep{\fill}} p{2.5cm}YYY|p{2.5cm}YYY YYY@{}}
\toprule
\multicolumn{4}{c|}{\textbf{AV}} & \multicolumn{7}{c}{\textbf{HV}} \\
Dataset & $a_{L}$ & $k_{L}$ & $\mathrm{R}^{2}_{L}$ &
Dataset & $a_{L}$ & $k_{L}$ & $\mathrm{R}^{2}_{L}$ & $a_{T}$ & $k_{T}$ & $\mathrm{R}^{2}_{T}$ \\
\midrule
Argoverse 2 & 0.110 & -0.809 & 0.864 &
CitySim-freeC & 0.910 & -0.388 & 0.963 & 0.310 & -0.660 & 0.741 \\
Waymo       & 4.111 & -0.230 & 0.999 &
CitySim-inA  & 2.911 & -0.304 & 0.997 & 7.912 & -0.267 & 0.954 \\
CATS ACC    & 7.311 & -0.112 & 0.992 &
CitySim-inB  & 81.926 & -0.015 & 0.997 & 0.610 & -0.475 & 0.975 \\
Ohio        & 1.510 & -0.381 & 0.987 &
highD Loc. 1 & 2.210 & -0.223 & 0.983 & 1.210 & -0.309 & 0.956 \\
MicroSimACC & 75.124 & -0.041 & 0.986 &
highD Loc. 2 & 1.510 & -0.420 & 0.987 & 2.210 & -0.371 & 0.993 \\
OA-Asta     & 6.411 & -0.113 & 0.995 &
highD Loc. 3 & 1.210 & -0.334 & 0.959 & 1.810 & -0.302 & 0.982 \\
OA-Vicolungo& 2.310 & -0.276 & 0.998 &
highD Loc. 4 & 0.710 & -0.464 & 0.970 & 1.810 & -0.311 & 0.988 \\
OA-Casale   & 3.611 & -0.198 & 0.988 &
highD Loc. 5 & 3.411 & -0.211 & 0.993 & 1.410 & -0.347 & 0.989 \\
OA-ZalaZone & 0.310 & -0.615 & 0.974 &
highD Loc. 6 & 0.510 & -0.573 & 0.950 & 3.111 & -0.253 & 0.991 \\
\midrule

\textbf{Average $\mathrm{R}^2_L$} &  &  & \textbf{0.976} &
\textbf{Average $\mathrm{R}^2_L$} &  &  & \textbf{0.972} &
\multicolumn{2}{c}{\textbf{Average $\mathrm{R}^2_T$}} & \textbf{0.972} \\
\textbf{Worst $\mathrm{R}^2_L$}   &  &  & \textbf{0.864} &
\textbf{Worst $\mathrm{R}^2_L$}   &  &  & \textbf{0.741} &
\multicolumn{2}{c}{\textbf{Worst $\mathrm{R}^2_T$}}   & \textbf{0.741} \\
\bottomrule
\end{tabular*}

\begin{tablenotes}[flushleft]
\footnotesize
\item Note: Subscripts $L$ and $T$ denote longitudinal and lateral, respectively. AV results are reported as longitudinal ($L$). OA denotes datasets from the OpenACC project (e.g., Asta, Vicolungo, Casale, ZalaZone).
\end{tablenotes}
\end{threeparttable}
\end{table}

\textbf{Fitting performance of the proposed shifted power law.} Table~\ref{tab:scaling_table} reports the complete fitting results of the proposed shifted power law across all AV and HV datasets. For AVs, all cases exhibit high coefficients of determination, with the average and minimum $\mathrm{R}^2$ values of $0.976$ and $0.864$, respectively. These high $\mathrm{R}^2$ values indicate that the overall fitting is accurate and robust, highlighting the generalizability of the power-law structure across different AV platforms and experimental conditions. Regarding HVs, the evaluation spans both naturalistic urban intersections (CitySim-freeC, CitySim-inA, CitySim-inB) and highway driving (highD across six locations, with both longitudinal and lateral behaviors). The model again demonstrates strong agreement with empirical data, achieving the average and minimum $\mathrm{R}^2$ values of $0.972$ and $0.741$, respectively. These $\mathrm{R}^2$ values remain very high, proving the proposed shifted power law well captures the dominant scaling trend. Nevertheless, they are slightly lower and more dispersed than those of AVs, revealing
more stochastic nature of human driving behavior. The fitted scale factor $a$ and decay exponent $k$ vary substantially across datasets, reflecting local driving styles and platform heterogeneity, while still maintaining the heavy-tailed scaling pattern. These results demonstrate that the shifted power law provides a uniform statistical representation of AV and HV dynamics across datasets, tasks (longitudinal vs.~lateral), and environments (urban vs.~highway). The model preserves consistency in normal driving while extending to rare events, supporting high-fidelity and safety-critical validation. We report more fitting results in~\ref{seca: fitting_results}.

\begin{figure}[!ht]
\centering
\includegraphics[width=1\linewidth]{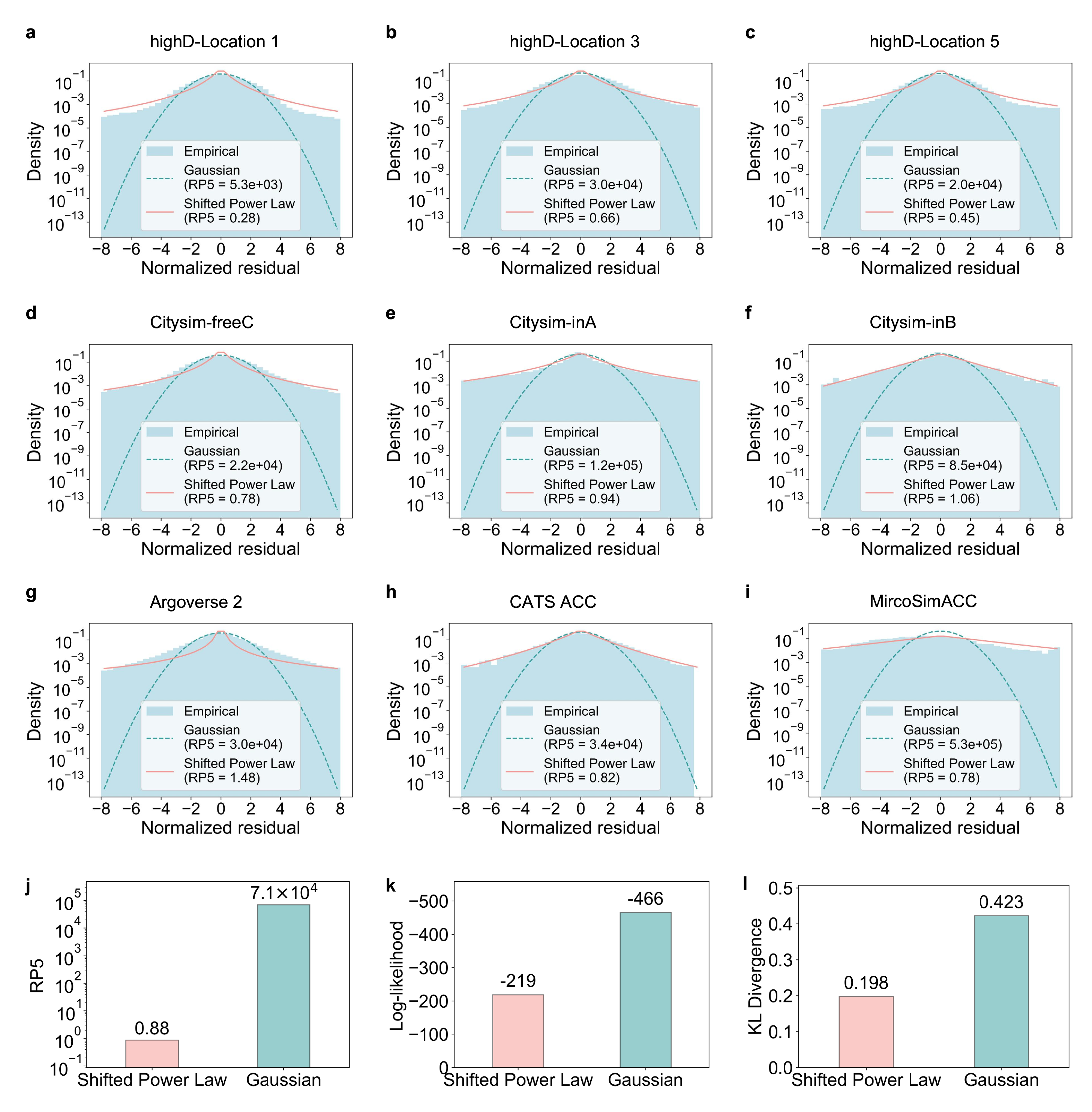}
\caption{\textbf{Comparison among the proposed shifted power law, the classic Gaussian distribution, and empirical distribution across AV and HV datasets.}
Tail accuracy is quantified by the RP5 metric, defined as the ratio of the empirical probability beyond five standard deviations ($|\bar{\sigma}|>5$) to the corresponding model-predicted probability. A value near 1.0 indicates accurate tail estimation.
\textbf{a–c}, Distributions of lateral normalized residual of three locations in the highD dataset.
\textbf{d–f}, Distributions of longitudinal normalized residual of the Citysim dataset.
\textbf{g–i}, Distributions of longitudinal normalized residual of three AV datasets: Argoverse~2, CATS ACC, and MicroSimACC.
\textbf{j–l}, Average values of RP5, Log-likelihood, and KL divergence achieved by the shifted power law and Guassian distribution across all datasets.}
\label{fig: compare}
\end{figure}

\textbf{Comparison with existing Gaussian distribution model.} To validate the efficacy of the proposed model, we conduct a comprehensive comparison with the classic Gaussian distribution baseline. Given that the accurate characterization of tail behavior is paramount for traffic safety, we first define a customized evaluation metric:  the \underline{R}atio of the \underline{P}robability of the normalized residual $|\bar{\sigma}| > \underline{5}$ estimated from observed data to the corresponding probability predicted by a model, short for ``RP5''. A RP5 value close to 1.0 indicates that the model accurately captures the likelihood of these rare, high-risk behavior. As shown in Figure~\ref{fig: compare}(a–i), the proposed shifted power law demonstrates exceptional tail alignment with the empirical distribution across all datasets. On the contrary, the Gaussian distribution consistently and significantly underestimates the probability of these extreme deviations. We report the detailed evaluation results in~\ref{seca: metrics}. Quantitatively, as depicted in Figure~\ref{fig: compare}(j), the average RP5 value for the Gaussian model is approximately $7.1 \times 10^{4}$, signifying an underestimation of risk by more than four orders of magnitude. The average RP5 value across all scenarios is approximately 0.88 for the shifted power law. This stark contrast indicates that the shifted power law provides a realistic assessment of tail risk, while the Gaussian model is fundamentally unsuitable for safety-critical applications.

We further compare the models using two metrics that evaluate the entire distribution: Log-Likelihood and Kullback-Leibler (KL) Divergence. The Log-Likelihood measures how well the model explains the entire empirical dataset, while the KL Divergence quantifies the overall similarity between the model and the empirical distribution. The detailed results, reported in~\ref{seca: metrics}, show that the shifted power law consistently outperforms the Gaussian distribution on both metrics. Specifically, as shown in Figure~\ref{fig: compare}(k–l), the average Log-Likelihood for the shifted power law and Gaussian across all datasets are -219 and -466, respectively, and the average KL Divergence are 0.198 and 0.423, respectively. This indicates that our model not only excels in capturing the critical long-tail behavior but also provides a more accurate and holistic fit to the entire spectrum of driving behaviors, from common safe maneuvers to rare risky events. The mathematical definitions of all metrics are introduced in Section~\ref{sec: metrics}.

\textbf{Comparison with established long-tail distributions.} While existing studies often default to a Gaussian assumption, we further investigate the nature of HV and AV driving behavior by benchmarking our model against two commonly used heavy-tailed distributions: the Laplace and Student’s $t$ distributions. The detailed results, reported in Section~\ref{seca: long_tail}, confirm our primary hypothesis. First, both the Laplace and Student’s $t$ distributions provide a significantly better fit to the empirical data than the Gaussian model, particularly in the tails. This conclusively substantiates that driving behavior distributions are inherently heavy-tailed. Second, and more importantly, our proposed shifted power law demonstrates a marked advantage over these heavy-tailed models, as the RP5 values achieved by the Laplace and Student’s $t$ distributions still deviate from the ideal by roughly an order of magnitude. This dual evidence not only substantiates that driving behavior distributions are heavy-tailed but also demonstrates that the shifted power law achieves superior fidelity relative to existing alternatives.

\subsection{Reflection of real-world driving behavior}\label{sec: 2.3}

\begin{figure}[!ht]
    \centering
    \includegraphics[width=1\linewidth]{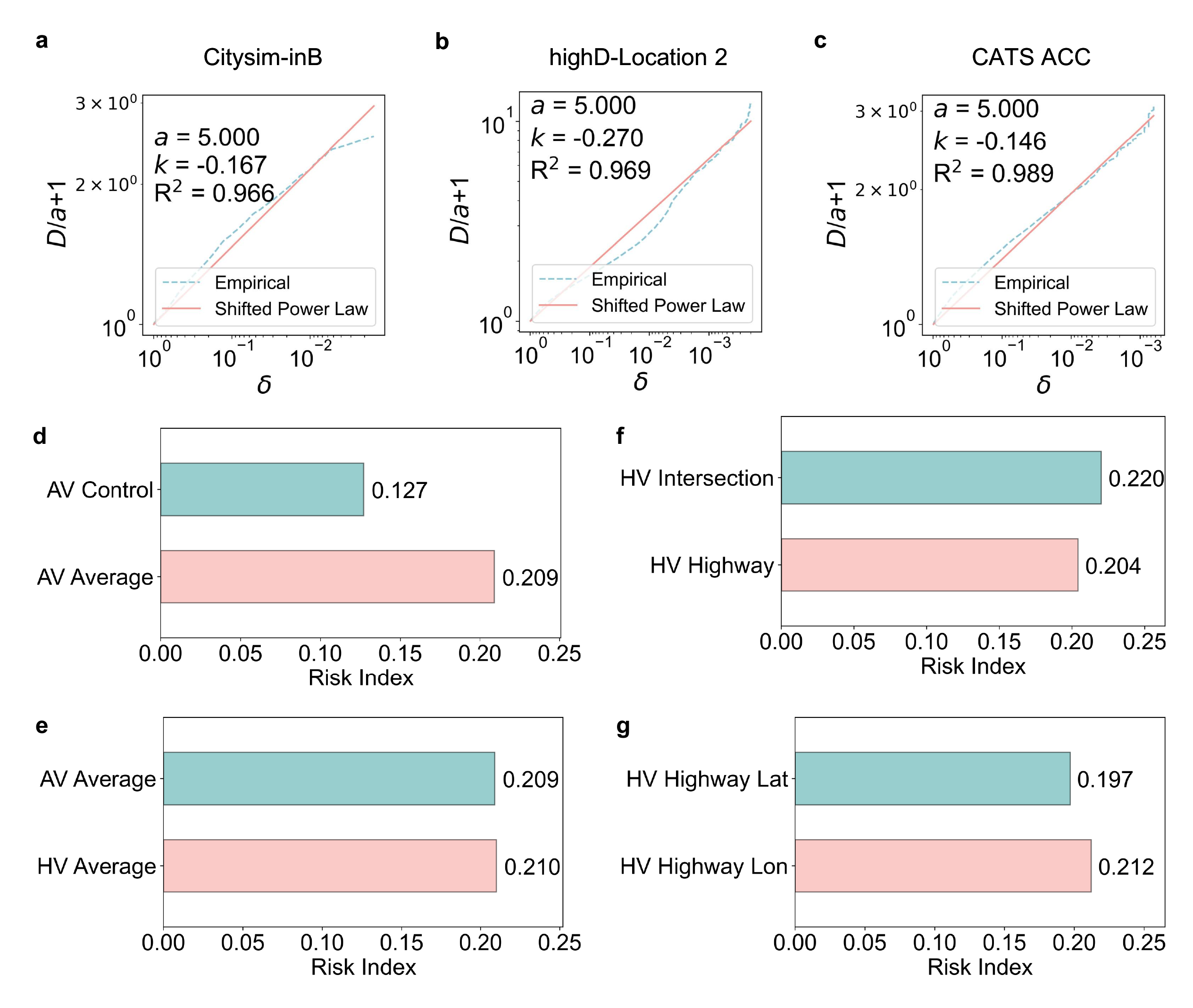}
    \caption{\textbf{Reflected real-world driving behavior with fixed scale at $a=5$.}
    \textbf{a–c}, Shifted power-law fitting under fixed scale across representative datasets including CitySim-inB (lateral), highD-Location~2 (longitudinal), and CATS ACC (longitudinal). 
    The model retains high fidelity to empirical distributions ($R^2 > 0.95$), confirming stability across different environments. Lon and Lat represent longitudinal and lateral directions, respectively.
    \textbf{d–g}, Comparison of the derived Risk Index ($|k|$) across vehicle types and scenarios. 
    \textbf{d}, Averaged Risk Index between AVs and HVs. 
    \textbf{e}, Controlled AV experiments showing improved predictability (smaller Risk Index). 
    \textbf{f}, HV driving at intersections and highways, where intersections exhibit higher stochasticity. 
    \textbf{g}, Comparison between HV lateral and longitudinal behaviors, showing higher risk in longitudinal control}
    \label{fig:reflect_a5}
\end{figure}

To further assess the robustness and generalizability of the shifted power law, we conducted an analysis with the scale parameter $a$ fixed to a value of 5. As shown in Figure~\ref{fig:reflect_a5}(a–d), the model maintains a strong agreement with the empirical data even under this constrained parameterization, with the $\mathrm{R}^2$ values across multiple datasets over 0.95, confirming that the shifted power law can consistently capture tail behaviors with only one parameter $k$. we present all fitting results in~\ref{seca: a5_results}.

Furthermore, with a fixed value of $a$, the decay exponent $k$ provides a powerful physical interpretation of driving behavior. We posit that the absolute value of $k$ denotes the predictable extent of a vehicle's actions in a given context; a smaller $|k|$ indicates a more predictable and less stochastic behavior, as the residual distribution exhibits a thinner tail. Consequently, we define $|k|$ as a Risk Index, which quantifies the inherent stochasticity and potential risk in driving behaviors. A higher Risk Index suggests a greater propensity for abrupt or extreme maneuvers, whereas a smaller value indicates smoother and more controllable motion patterns.

To validate the effectiveness of the shifted power law and its derived Risk Index in reflecting fundamental driving behavior, we analyze data from a controlled experiment \cite{ma2025real}. In this experiment, the following AV was governed by a custom, deterministic linear controller, a simple algorithm that computed acceleration as a direct function of ego speed and headway. This hand-tuned controller was designed for high stability, resulting in exceptionally low behavioral variability. As shown in Figure~\ref{fig:reflect_a5}(d), the analysis of this controlled data yields a Risk Index of 0.127, which is significantly smaller than those observed in the other normal driving scenarios (with an average of 0.209). The result serves as a critical validation point: the shifted power law framework successfully differentiates this highly stable, algorithmically controlled behavior from the more stochastic behaviors of the previously discussed AV systems. By cleanly capturing this contrast, the model demonstrates its sensitivity and robustness in quantifying the predictable extent of real-world driving behavior.

We further compare the driving behavior across various scenarios using the risk index, as shown in Figures~\ref{fig:reflect_a5}(e–g). Specifically, Figures~\ref{fig:reflect_a5}(e) show that the overall average risk indexes for HVs and AVs are nearly identical ($0.210$ vs.\ $0.209$), suggesting that AVs can achieve comparable predictability with HVs under normal driving conditions \footnote{We only compare the longitudinal driving behaviors between AVs and HVs}. This is probably because many of these AV controls are designed with imitation learning based on human driver data.
For HVs, the comparison between intersections and highways (Figure~\ref{fig:reflect_a5}(f)) reveals that intersection driving produces a slightly greater Risk Index ($0.220$ vs.\ $0.204$), implying a higher degree of uncertainty at intersections due to more complex interactions and conflicts. Similarly, Figure~\ref{fig:reflect_a5}(g) shows that longitudinal driving generally exhibits greater stochasticity than lateral behavior ($0.212$ vs.\ $0.197$), indicating that the longitudinal driving behaviors are more stochastic than lateral ones on highways, which may induce more traffic crashes (please refer to Section~\ref{sec: 2.4} for further analysis). These results suggest that the fixed-scale shifted power law not only preserves fitting accuracy but also enables a meaningful, interpretable measure of real-world driving risk, providing a valuable insight into measuring stochastic driving behaviors of both HVs and AVs.


\subsection{Crash rate comparison and validation in simulation}
\label{sec: 2.4}

\begin{figure}[!ht]
\centering
\includegraphics[width=1\linewidth]{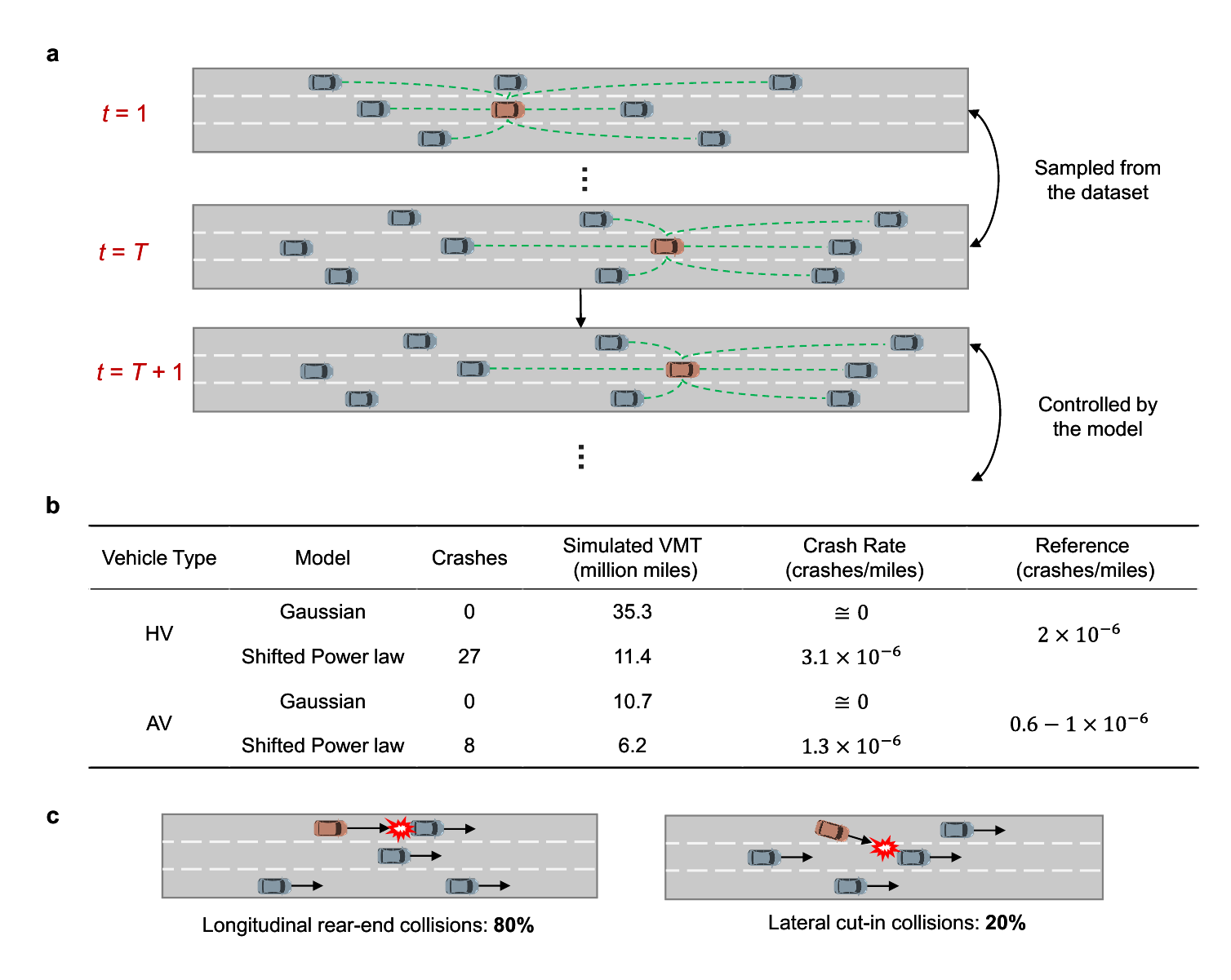}
\caption{\textbf{Crash rate comparison and validation through large-scale simulation.}
\textbf{a}, Overview of the agent–based simulation framework where each vehicle interacts with at most eight surrounding vehicles. The initial state ($t = 1, \cdots, T$) is sampled from the dataset (HV from the highD dataset and AV from the Waymo dataset), and the movements of all vehicles in the next steps ($t=T+1,\cdots$) are controlled by the model (Gaussian or shifted power law).
\textbf{b}, Aggregated crash outcomes and empirical benchmarks, showing that the shifted power law model reproduces real-world crash rates for HVs/AVs. 
\textbf{c}, Directional breakdown of simulated crashes across longitudinal and lateral dimensions, revealing that approximately 80\% of crashes are rear-end impacts, consistent with empirical Risk Index asymmetry and naturalistic driving data.
}
\label{fig:fig5}
\end{figure}

To quantitatively evaluate the safety implications of behavior modeling under uncertainty, we conduct large-scale closed-loop simulations using a proprietary agent-based traffic environment. As illustrated in Figure~\ref{fig:fig5}(a), our developed platform simulates traffic flow where each vehicle, governed by a specified behavioral model (i.e., the Gaussian or shifted power law), dynamically interacts with up to eight surrounding vehicles. The simulation is initialized by sampling a sequence of historical states (from $t=1$ to $t=T$) from real-world trajectory datasets. For each subsequent time step, the model predicts the acceleration distribution for every vehicle; a concrete acceleration value is then sampled from this distribution to update the vehicle's kinematic state. This process iterates recursively, using the newly generated states (e.g., $t=2,3,\cdots,t=T+1$) to predict subsequent states ($t=T+2,\cdots$), thereby propagating the simulation forward in time. We examine two distinct operational scenarios—HVs simulated with data from the highD dataset and AVs simulated with data from the Waymo dataset—to compare the crash rates generated by the two candidate models. A comprehensive description of the simulation is provided in Section~\ref{sec: Simulation}.

The simulated crash rates for each model and vehicle type are summarized in Figure~\ref{fig:fig5}(b). For HVs, the Gaussian model resulted in zero crashes over 35.3 million miles, a rate which, when statistically compared to the real-world baseline of around $2\times10^{-6}$~\cite{nhtsa2024report} using the Z-test detailed in Section~\ref{sec: z_test}, is significantly lower ($z = -8.40$, $p < 0.001$). On the contrary, the shifted power law model yields a crash rate of $3.1\times10^{-6}$, which is not statistically different from the baseline ($z = 0.88$, $p > 0.3$). A parallel analysis for AVs against a reference range of $0.6$–$1.0\times10^{-6}$~\cite{abdel2024matched,kalra2016driving} reveals a similar trend. The Gaussian model again significantly underestimates the crash rate ($z = -5.16$ for the upper bound, $p < 0.001$), while the shifted power law model's rate of $1.3\times10^{-6}$ is not a statistically significant deviation ($z = 0.75$, $p > 0.4$). 

Importantly, the relative relationship between HV and AV crash rates is also correctly reproduced: the simulated AV crash rate is significantly lower than that of HVs, consistent with large-scale empirical findings that autonomous systems generally experience fewer collisions per mile than human drivers~\cite{abdel2024matched}. This alignment between model outcomes and real-world statistics emerges naturally from the unified stochastic law, without any empirical correction, underscoring the model’s explanatory power. In addition, as shown in Figure~\ref{fig:fig5}(c), the crash type distribution for the highD dataset shows longitudinal rear-end collisions: 80\% and lateral cut-in collisions: 20\%, which is consistent with the risk index calculated in Section \ref{sec: 2.3} (longitudinal $>$ lateral), as well as large-scale studies of naturalistic and autonomous driving datasets~\cite{liu2023safety,fridman2019autonomous,iihs2022rearend}. These results collectively indicate that the shifted power law model produces crash rates that are statistically consistent with real-world data, whereas the Gaussian model consistently and significantly underestimates crash risk.

\section{Discussion}\label{sec: 3}

\textbf{Generality of stochastic scaling in driving behavior.}  
This study identifies a simple yet universal statistical law that governs stochastic driving behaviors across both human-driven and autonomous systems. The shifted power law consistently fits residual accelerations from diverse datasets and geographic contexts, maintaining accuracy even when the scale parameter is fixed. These results demonstrate that heavy-tailed variability is not an artifact of sampling or modeling assumptions, but rather a stable signature of how drivers—human or machine—regulate control under uncertainty. Such universality indicates that, despite technological or contextual differences, both cognitive and algorithmic drivers follow a shared probabilistic structure shaped by physical constraints, perceptual delay, and adaptive control mechanisms.

\textbf{Interpretable quantification of risk.}  
The decay exponent $k$, represented by its absolute value $|k|$, defines a Risk Index that links behavioral variability to control difficulty. Smaller $|k|$ values reflect predictable and smooth motion, whereas larger ones correspond to abrupt and less controllable actions. This unified measure correctly positions deterministic AV controllers at the low-risk end and places naturalistic human driving at higher variability levels, illustrating that the same statistical descriptor spans engineered and human control regimes. The Risk Index further reveals intrinsic anisotropy: longitudinal behaviors exhibit higher stochasticity than lateral behaviors, and intersections show greater variability than highways. These findings align with the physical limits of vehicle dynamics and the cognitive demands of speed regulation and gap management, offering a compact and interpretable metric of inherent driving risk across diverse agents and environments.

\textbf{From behavioral statistics to safety outcomes.}  
Embedding the shifted power law within an agent-based interaction-aware simulator bridges statistical realism with operational safety. When residuals are sampled from the shifted power law distribution rather than Gaussian noise, the simulator reproduces rare yet consequential events, yielding crash frequencies that align closely with real-world statistics. Directional asymmetry in the Risk Index also manifests physically: locations with higher longitudinal variability produce a dominance of rear-end collisions, consistent with highway crash patterns. This correspondence confirms that the statistical anisotropy inferred from data translates into tangible safety outcomes, validating the shifted power law as an effective bridge between behavioral modeling and real-world risk.

\textbf{Implications for simulation, validation, and system design.}  
These findings underscore the necessity of rare-event fidelity in behavioral modeling. Conventional Gaussian or truncated models may achieve low average error but systematically underrepresent extreme maneuvers that dominate safety outcomes. The shifted power law preserves these rare tails, enabling realistic edge-case generation and interpretable simulation of risk. The derived Risk Index provides a portable calibration target: by lowering $|k|$ in high-risk regimes, developers can directly associate controller adjustments with measurable safety improvements. Furthermore, reporting $(a, k)$ and the corresponding Risk Index across scenario categories can standardize behavioral variability profiling for both HVs and AVs, facilitating digital-twin validation, cross-platform benchmarking, and regulatory auditing.

\textbf{Scalability, limitations, and outlook.} Methodologically, this framework decouples the estimation of mean motion from the modeling of residual variability, ensuring both predictive accuracy and statistical realism. While the present analysis focuses on short-horizon acceleration dynamics, future extensions could explore coupled longitudinal–lateral motions, jerk, and yaw-rate constraints to capture richer control dependencies. Cross-cultural and cross-environmental calibration of $(a, k)$ using federated or privacy-preserving learning may further reveal how socio-technical factors shape stochastic scaling across cities and driving styles. Another promising direction lies in coupling the Risk Index with online adaptive controllers, where real-time estimates of $|k|$ can inform proactive adjustments of headway, speed, or aggressiveness under changing risk levels. Finally, integrating rare-event acceleration techniques such as importance sampling or adversarial scenario generation guided by the tail exponent $k$ can provide statistically grounded yet computationally efficient validation pipelines. Despite these considerations, the results outline a coherent path forward, a universal scaling law describes behavioral variability, an interpretable index quantifies scenario risk, and a rare-event faithful generator reproduces safety outcomes. Together, these components establish a unified foundation for comparing, stress-testing, and improving human and autonomous driving systems on a shared statistical basis.

\section{Methods}\label{sec: 4}
\subsection{Datasets}\label{sec: 4.1}

To ensure the robustness and generalizability of our findings, we employed large-scale, multi-source trajectory datasets that capture both AV and HV behaviors across diverse driving environments. The datasets include \textit{highD} (Germany, HV) \cite{krajewski2018highd}, \textit{CitySim} (USA, China, HV) \cite{zheng2024citysim}, \textit{MicroSimACC} (USA, AV) \cite{yang2024microsimacc}, \textit{CATS} (USA, AV) \cite{shi2021empirical,ma2025real}, \textit{OpenACC} (Europe, AV) \cite{makridis2021openacc}, \textit{Central Ohio} (USA, AV) \cite{xia2023automated}, \textit{Waymo Open} (USA, AV) \cite{sun2020scalability,hu2022processing}, and \textit{Argoverse~2} (USA, AV) \cite{wilson2023argoverse}. These datasets collectively encompass both naturalistic and controlled experiments across highway, suburban, and urban settings, supporting comprehensive analyses of driving behaviors and controller performance. A detailed description of each dataset, including collection settings, data structures, and sampling rates, is provided in \ref{sec:dataset_details}.

\subsection{Driving behavior prediction}\label{sec: prediction}

We develop machine learning models based on Long Short-Term Memory (LSTM) networks~\cite{huang2020probabilistic} to predict the mean value $\hat{y}_{T+1}$ and standard deviation $\hat{\gamma}_{T+1}$ of the ego vehicle's acceleration at time $T+1$ based on the previous state $x_{1:T}$. Specifically, the state $x_{1:T}$ is first input to the LSTM and then processed by two individual feed-forward networks (FFNs) to predict $\hat{y}_{T+1}$ and $\hat{\gamma}_{T+1}$, respectively.
The state for AVs includes the ego and leading vehicle’s acceleration, speed, and bumper-to-bumper distance over the past $T$ time points.
For HVs, we consider two-dimensional (2D) predictions and incorporate richer context, including the velocity and acceleration of the ego and its surrounding vehicles, their relative distances, lane ID, etc. Please refer to \ref{seca: prediction} for details.

In this study, we adopt a single-step prediction paradigm with a temporal discretization of $\Delta t = 0.2$ s and a historical context of $T=12$ steps, resulting in a 0.2 s prediction horizon. The investigation into varying prediction horizons, detailed in~\ref{seca: prediction_horizon}, revealed that the distributions of the normalized residual for longer horizons (0.4\,s, 0.6\,s, and 0.8\,s) exhibit statistically similar trends to the 0.2 s baseline, indicating that the model's predictive uncertainty scales consistently and validating the shorter horizon as a reliable and efficient indicator of model behavior. Furthermore, the comparative analysis between single-step and multi-step prediction paradigms in~\ref{seca: multi_step_prediction} demonstrates that both modes share similar trends in their normalized residual distributions across equivalent horizons, which suggests that the model's uncertainty estimates are well-calibrated.

\subsection{Derivation of distribution}\label{sec: derivation}

This section derives the analytical expressions of the cumulative distribution function (CDF) and probability density function (PDF) for the shifted power-law distribution~$\mathcal{P}$ of the normalized residual~$\bar{\sigma}$. 
We start from the fundamental relationship between the residual magnitude~$|\sigma|$ and its violation probability~$\delta(\sigma)$ defined in Equations~\eqref{eq:violation_rate} and~\eqref{eq:shifted_power_law}. 
Solving this relationship for the tail probability yields:
\begin{equation}
    \delta(\sigma) = \left( 1 + \frac{|\sigma|}{a} \right)^{1/k}.
\end{equation}
Assuming that $\bar{\sigma}$ follows a symmetric, unimodal distribution centered at zero (consistent with the observed balance between acceleration and deceleration behaviors), 
we recover the signed distribution from the absolute-value relationship above. 
For non-negative arguments $\sigma \ge 0$, by definition, the tail probability relates to the CDF of the absolute value $F^{\mathcal{P}}$ through $\frac{1}{2}\delta(\sigma) = 1 - F^{\mathcal{P}}(|\sigma|)$. 
Substituting $\delta(\sigma)$ gives:
\begin{equation}
    F^{\mathcal{P}}(\sigma) = 1 - \frac{1}{2}\delta(\sigma) 
    = 1 - \frac{1}{2} \left( 1 + \frac{\sigma}{a} \right)^{1/k}, 
    \quad \sigma \ge 0.
\end{equation}
For negative arguments $\sigma < 0$, symmetry dictates that the CDF equals half the probability that the magnitude exceeds~$|\sigma|$:
\begin{equation}
    F^{\mathcal{P}}(\sigma) 
    = \frac{1}{2}\delta(\sigma) 
    = \frac{1}{2} \left( 1 - \frac{\sigma}{a} \right)^{1/k}, 
    \quad \sigma < 0.
\end{equation}
Combining these two cases yields the complete, continuous CDF:
\begin{equation}
F^{\mathcal{P}}(\sigma) =
\begin{cases}
\frac{1}{2} \left( 1 - \dfrac{\sigma}{a} \right)^{\frac{1}{k}}, & \sigma < 0,\\[4mm]
1 - \dfrac{1}{2} \left( 1 + \dfrac{\sigma}{a} \right)^{\frac{1}{k}}, & \sigma \ge 0.
\end{cases}
\end{equation}
This CDF is symmetric about~$\sigma = 0$ and satisfies the logical boundary conditions $F^{\mathcal{P}}(0)=0.5$, $\lim_{\sigma\to-\infty}F^{\mathcal{P}}(\sigma)=0$, and $\lim_{\sigma\to+\infty}F^{\mathcal{P}}(\sigma)=1$. 

The CDF ensures that the total probability mass integrates to one and serves as the foundation for the analytical form of the PDF used in the main text. The corresponding PDF is obtained by differentiating the CDF with respect to~$\sigma$. 
For $\sigma>0$,
\begin{equation}
    f^{\mathcal{P}}(\sigma) 
    = \frac{d}{d\sigma} 
      \left[ 1 - \frac{1}{2} \left( 1 + \frac{\sigma}{a} \right)^{1/k} \right] 
    = -\frac{1}{2ak} 
      \left( 1 + \frac{\sigma}{a} \right)^{1/k - 1}.
\end{equation}
For $\sigma<0$, similarly,
\begin{equation}
    f^{\mathcal{P}}(\sigma) 
    = \frac{d}{d\sigma} 
      \left[ \frac{1}{2} \left( 1 - \frac{\sigma}{a} \right)^{1/k} \right] 
    = -\frac{1}{2ak} 
      \left( 1 - \frac{\sigma}{a} \right)^{1/k - 1}.
\end{equation}
Consequently, the unified expression of the PDF for all real~$\sigma$ is
\begin{equation}
f^{\mathcal{P}}(\sigma) 
= -\frac{1}{2ak} 
  \left( 1 + \frac{|\sigma|}{a} \right)^{\frac{1}{k} - 1}, 
  \quad \forall \sigma \in \mathbb{R}.
\end{equation}
This formulation defines a symmetric, heavy-tailed probability distribution that satisfies 
\[
\int_{-\infty}^{+\infty} f^{\mathcal{P}}(\sigma)\, d\sigma = 1,
\]
for all $a > 0$ and $k < 0$. 
As $|\sigma| \to \infty$, the tail asymptotically follows 
$f^{\mathcal{P}}(\sigma) \propto |\sigma|^{1/k - 1}$, 
revealing that smaller (more negative) values of $k$ produce heavier tails. These properties underpin the physical interpretation of $k$ as the decay exponent controlling behavioral unpredictability, and provide the analytical foundation for the results presented in Section~\ref{sec: 2.1}.

\subsection{Evaluation metrics}\label{sec: metrics}

To quantitatively evaluate how well each candidate distribution reproduces the empirical characteristics of driving behavior, we employ three complementary metrics that jointly assess overall goodness-of-fit and, crucially, tail fidelity. First, we introduce the RP5, a custom metric designed to measure a model's accuracy in capturing long-tail behavior, defined as the ratio of the empirical probability mass to the model's probability mass in the extreme region where $|\sigma| \geq 5$:
\begin{equation}
\mathrm{RP5} = \frac{\int_{|\sigma|\geq 5} f^{\mathrm{empirical}}(\sigma) \mathrm{d}\sigma}{\int_{|\sigma|\geq 5} f^{\mathrm{model}}(\sigma) \mathrm{d}\sigma},
\label{eq:RP5}
\end{equation}
where $f^{\mathrm{empirical}}(\sigma)$ and $f^{\mathrm{model}}(\sigma)$ denote the PDF presented by the empirical data and approximated by the model, respectively. A RP5 value close to 1 indicates that the fitted model accurately matches the empirical tail distribution, whereas values deviating from 1 reflect underestimation or overestimation of rare yet safety-critical events. In addition, we use the standard Log-Likelihood to measure the overall goodness-of-fit across the entire distribution:
\begin{equation}
\text{Log-likelihood} = \int_{\sigma} \log \, f^{\mathrm{model}}(\sigma) \mathrm{d}\sigma,
\label{eq:loglikelihood}
\end{equation}
where a higher value signifies a better match. Also, we compute the Kullback–Leibler (KL) Divergence to evaluate the information-theoretic similarity between the empirical data and the model, as follows:
\begin{equation}
\mathrm{KL}\big(f^{\mathrm{empirical}} \,\|\, f^{\mathrm{model}}\big) = \int_{\sigma} f^{\mathrm{empirical}}(\sigma) \log \frac{f^{\mathrm{empirical}}(\sigma)}{f^{\mathrm{model}}(\sigma)} \mathrm{d}\sigma.
\label{eq:KL}
\end{equation}
Smaller KL divergence values indicate closer agreement between the model and empirical data, with a value of zero denoting a perfect match. Together, these three metrics jointly assess (i) the accuracy of rare-event tail modeling (RP5), (ii) the overall likelihood of observed data (log-likelihood), and (iii) the information-theoretic similarity between distributions (KL divergence).

\subsection{Crash rate validation via Negative Binomial modeling}\label{sec: z_test}

To evaluate whether the crash rate simulated by a given behavior model (i.e., the proposed shifted power law or Gaussian distribution) statistically matches a known real-world baseline, we adopt a hypothesis testing framework based on the Negative Binomial (NB) model \cite{denwood2019hypothesis}. Let $p$ denote the true crash rate (e.g., average crashes per million VMT) obtained from real-world driving data. For each candidate behavior model, we record the number of crashes $n_i$ ($i \in \{1,2\}$) occurring within $m_i$ million miles of simulation, and compute the estimated rate $p_i = \frac{n_i}{m_i}$. We then test the null hypothesis:
\begin{itemize}
    \item $H_0$: The model’s observed crash rate equals the true rate, i.e., $p_i = p$
    \item $H_1$: The model’s crash rate deviates from the true rate, i.e., $p_i \neq p$
\end{itemize}
The statistical significance is evaluated using a standard Z-test:
\begin{equation}
z_i = \frac{p_i - p}{\sqrt{\frac{p(1 - p)}{m_i}}},
\end{equation}
where $z_i$ approximately follows a standard normal distribution under $H_0$. A result of $|z_i| > 1.96$ implies statistical rejection at the 95\% confidence level.

This test allows us to rigorously quantify whether a behavior model not only simulates plausible trajectories but also replicates real-world safety outcomes. In our experiments, the power law model’s simulated crash rate yields $|z| < 1.96$, showing statistical consistency with empirical crash rates, while the Gaussian model overestimates crashes and is rejected under the same test.

\subsection{Simulation environment}\label{sec: Simulation}

To evaluate how the learned stochastic behavior models affect safety-critical outcomes, we build a simulation framework that replicates the geometric layout and traffic flow characteristics of the highD dataset (for HV simulations) and Waymo dataset (for AV simulations). For each scenario, initial vehicle positions, velocities, and inter-vehicle distances are sampled from recorded trajectories. The trained model 
(see Section~\ref{sec: prediction}) generates acceleration estimates $\hat{y}_{T+1}$ and corresponding deviations $\hat{\sigma}_{T+1}$ based on the sampled initial state, which are used to reconstruct the true distribution of the vehicle's acceleration according to Equation~\eqref{eq: y_tilde}. Subsequently, we sample stochastic accelerations from the reconstructed distribution. These sampled accelerations are recursively applied to update each vehicle’s state at each simulation step. At the end of each simulation step, we update the surrounding vehicles of each vehicle and detect collisions among them.

For simulations of AV/HV, we consider two behavior prediction models under identical conditions: a baseline Gaussian model and the proposed shifted power law with the PDF of the normalized residual as $\mathcal{N}(0,1)$ and Equation~\eqref{eq:scaling_pdf}, respectively. For each scenario, we simulate millions of vehicle miles travelled (VMT), ensuring robust statistical evaluation across diverse traffic interactions. This setup enables a direct comparison of safety implications between the Gaussian and shifted power law models under identical environmental and control conditions.

To implement the abovementioned simulations, we develop a high-fidelity agent-based simulation environment. In this simulator, each vehicle is instantiated as an autonomous agent that continuously records its full state, including acceleration, velocity, location, Vehicle Miles Traveled (VMT), the dynamic states of its surrounding vehicles, etc. This design allows for a holistic capture of traffic micro-dynamics. A key feature of our simulator is its seamless integration with trained prediction models, enabling them to serve as the behavioral core for each agent. The framework is capable of simulating realistic driving environments by incorporating physical parameters such as vehicle dimensions, type, and road geometry. For the experiments in this study, all vehicles are controlled directly by the trained prediction model, demonstrating its capability to function independently without reliance on an ensemble of models or heuristic fallbacks. Detailed simulation configurations are provided in ~\ref{seca: simulation}.

\section*{Data availability}

The datasets used to support the findings of this study are publicly available from their respective sources. 
The \href{https://www.highD-dataset.com}{\textit{highD}} dataset, which captures naturalistic highway driving behavior in Germany using drone recordings, is openly accessible. 
The \href{https://github.com/daibi/CitySim}{\textit{CitySim}} dataset, containing urban vehicle trajectories recorded by drones across diverse U.S.\ traffic scenarios, is also available.

\noindent
Multiple AV experiments worldwide were utilized, including 
\href{https://github.com/microSIM-ACC/ICE}{\textit{MicroSimACC}}, 
\href{https://github.com/CATS-Lab/Filed-Experiment-Data-ACC_Data}{\textit{CATS ACC}}, \href{https://github.com/MarkMaaaaa/CATS-UWMadison-AV-Data}{\textit{CATS UWM}},
\href{https://data.europa.eu/data/datasets/9702c950-c80f-4d2f-982f-44d06ea0009f?locale=en}{\textit{OpenACC}}, 
\href{https://catalog.data.gov/dataset/advanced-driver-assistance-system-adas-equipped-single-vehicle-data-for-central-ohio}{\textit{Central Ohio (single-vehicle ADAS)}} and 
\href{https://catalog.data.gov/dataset/advanced-driver-assistance-system-adas-equipped-two-vehicle-data-for-central-ohio}{\textit{Central Ohio (two-vehicle ADAS)}} 
\href{https://waymo.com/open/}{\textit{Waymo Open}}, 
and \href{https://www.argoverse.org/av2.html}{\textit{Argoverse~2}}.

\section*{Code availability}
The source codes used in this study are publicly available at \url{https://github.com/CATS-Lab/Shifted_Power_Law}.

\newpage


\printbibliography

\newpage

\section*{Acknowledgement}

We gratefully acknowledge Prof. Henry Liu and his team at the University of Michigan, Ann Arbor, whose work on \textit{Learning naturalistic driving environment with statistical realism} provided key inspiration and empirical support for this study. We also thank Dr. Richard Cheng and his team at the California Institute of Technology for their work on \textit{Limits of Probabilistic Safety Guarantees when Considering Human Uncertainty}, which offered valuable theoretical insights and comparative references for our analysis. 
This research was supported by grants from the National Natural Science Foundation CPS: Small: Cyber-Physical Phases of Mixed Traffic with Modular \& AVs: Dynamics, Impacts and Management (No. 2313578).

\section*{Author contributions}
Wang Chen: Methodology, Formal analysis, Software, Writing - original draft. Heye Huang: Methodology, Formal analysis, Writing - original draft.  Ke Ma: Formal analysis, Writing - review \& editing.  Hangyu Li: Software. Shixiao Liang: Writing - original draft. Hang Zhou: Writing - original draft. Xiaopeng Li: Conceptualization, Supervision, Writing - review \& editing.

\section*{Competing interests}
The authors declare no competing interests.

\newpage

\begin{center}
    \begin{tabular}{c}
     \\
     \\
     \\
     \\
     \\
     \\
     \\
     \\
    \textbf{Supplementary Information}  \\
    \\
    \\
    \\

    \textbf{Unveiling Uniform Shifted Power Law} \\
    \textbf{in Stochastic Human and Autonomous Driving Behavior} \\
    \\
    
    Wang Chen$^{a,\dag}$, Heye Huang$^{a,\dag}$, Ke Ma$^a$, Hangyu Li$^a$, Shixiao Liang$^a$, Hang Zhou$^a$, Xiaopeng Li$^{a,*}$ \\
    \\
    
    $^a$ Department of Civil and Environmental Engineering, University of Wisconsin–Madison, \\
    Madison, Wisconsin, 53706, USA\\
    $^*$ Corresponding Author, xli2485@wisc.edu\\
    $\dag$ These authors contributed equally to this work.\\

    \end{tabular}
\end{center}

\appendix
\setcounter{figure}{0}
\setcounter{table}{0}
\renewcommand{\thefigure}{S\arabic{figure}}
\renewcommand{\thetable}{S\arabic{table}}


\newpage
\section{Supplementary figures}

\begin{figure}[!ht]
    \centering
    \includegraphics[width=0.8\linewidth]{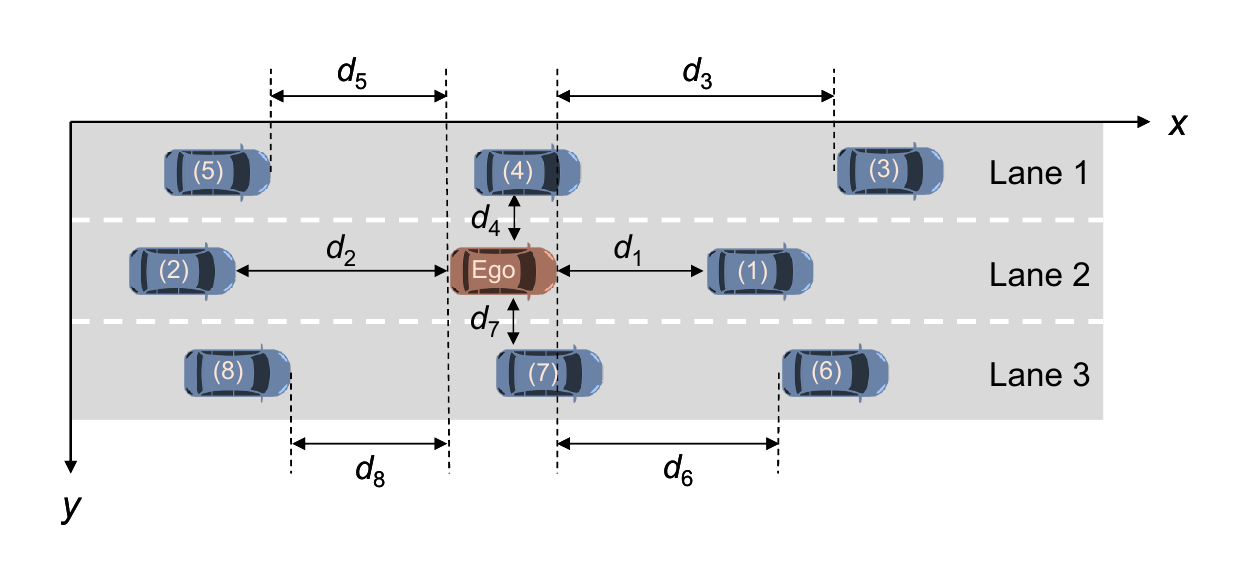}
    \caption{\textbf{Illustration of the state for predicting the acceleration of the ego vehicle.} Eight surrounding vehicles are considered when predicting the acceleration of the ego vehicle, including (1) the proceeding vehicle, (2) the following vehicle, (3) the left proceeding vehicle, (4) the left alongside vehicle, (5) the left following vehicle, (6) the right proceeding vehicle, (7) the right alongside vehicle, and (8) the right following vehicle. Also, the corresponding distances between the ego vehicle and its surrounding vehicles ($d_1$–$d_8$) and the lane id are considered.}
    \label{figa: prediction_state}
\end{figure}

\newpage
\begin{figure}[!ht]
    \centering
    \includegraphics[width=0.8\linewidth]{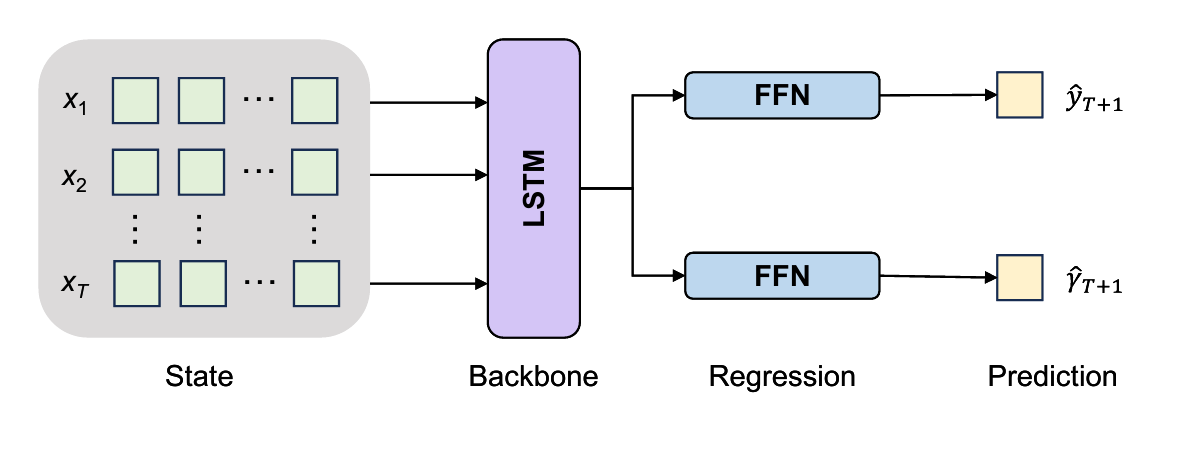}
    \caption{\textbf{An overview of one-step prediction.} The state ($x_{1:T}$) is first input into the backbone, i.e., the LSTM network, which then processes it through two individual FFNs to predict the mean value ($\hat{y}_{T+1}$) and standard deviation ($\hat{\gamma}_{T+1}$). The state includes the acceleration, speed, relative distance, lane id, etc., of the ego and its surrounding vehicles. LSTM and FFN denote long short-term memory and feed-forward neural network, respectively.}
    \label{figa: predict_network}
\end{figure}


\newpage
\begin{figure}[!ht]
\centering
\includegraphics[width=1\linewidth]{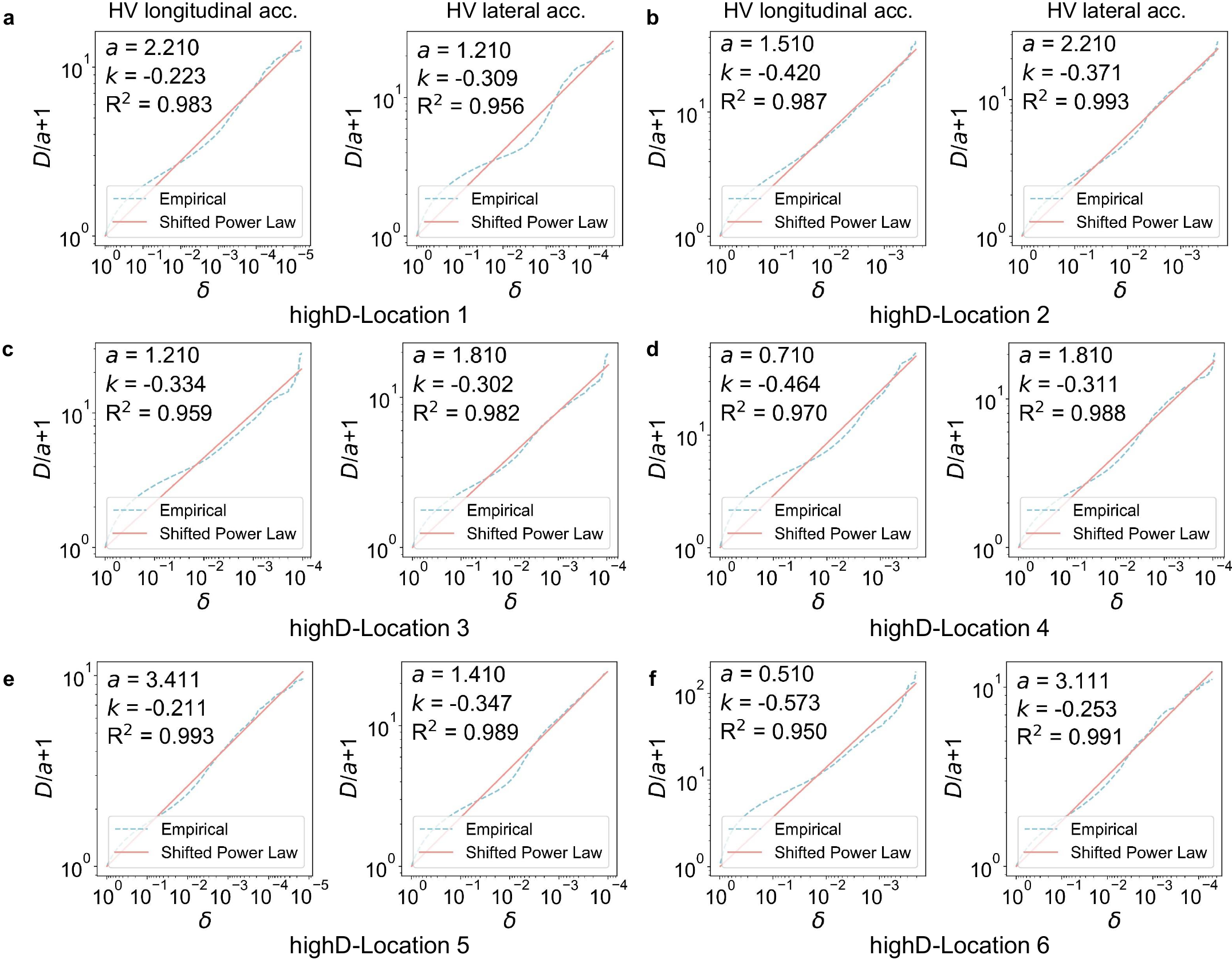}
\caption{\textbf{Fitting results of HVs on the highD dataset using the shifted power law.}
\textbf{a–f}, HV longitudinal and lateral acceleration distributions across six highway locations (Location~1–6) in Germany. 
The shifted power law (red) exhibits high fidelity to empirical data (blue), with fitted parameters $a$, $k$, and $\mathrm{R}^2$ indicating robust tail representation and strong generalization across highway environments.}
\label{fig:figC11}
\end{figure}

\newpage
\begin{figure}[!ht]
\centering
\includegraphics[width=1\linewidth]{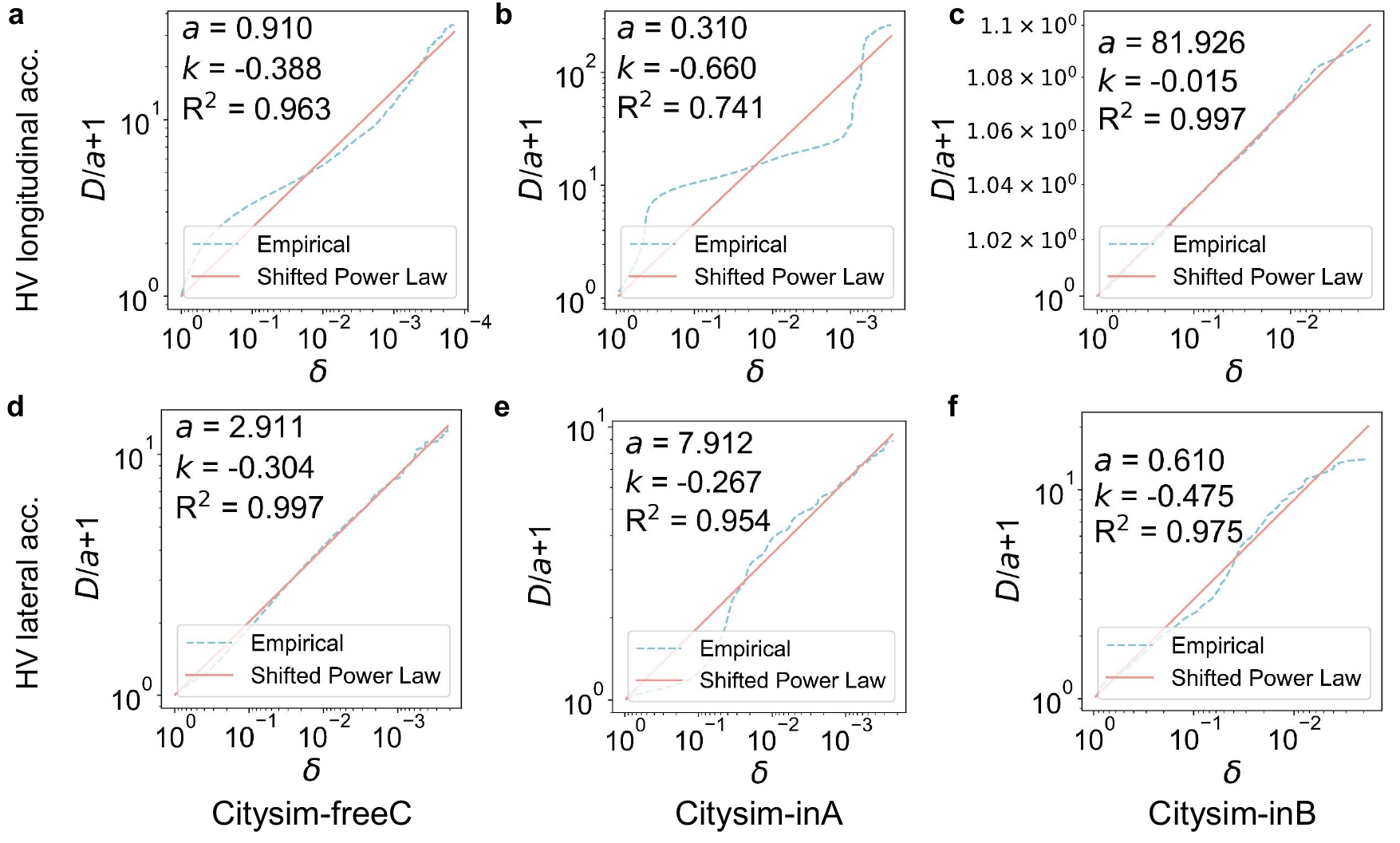}
\caption{\textbf{Fitting results of HVs on the CitySim dataset.}
\textbf{a–f}, HV longitudinal and lateral acceleration distributions from CitySim-freeC (China), CitySim-inA (non-signalized, U.S.), and CitySim-inB (signalized, U.S.) scenarios. 
The shifted power law (red) closely matches the empirical distributions (blue), capturing stochastic interactions under diverse urban traffic conditions.}
\label{fig:figC12}
\end{figure}

\newpage
\begin{figure}[!ht]
\centering
\includegraphics[width=1\linewidth]{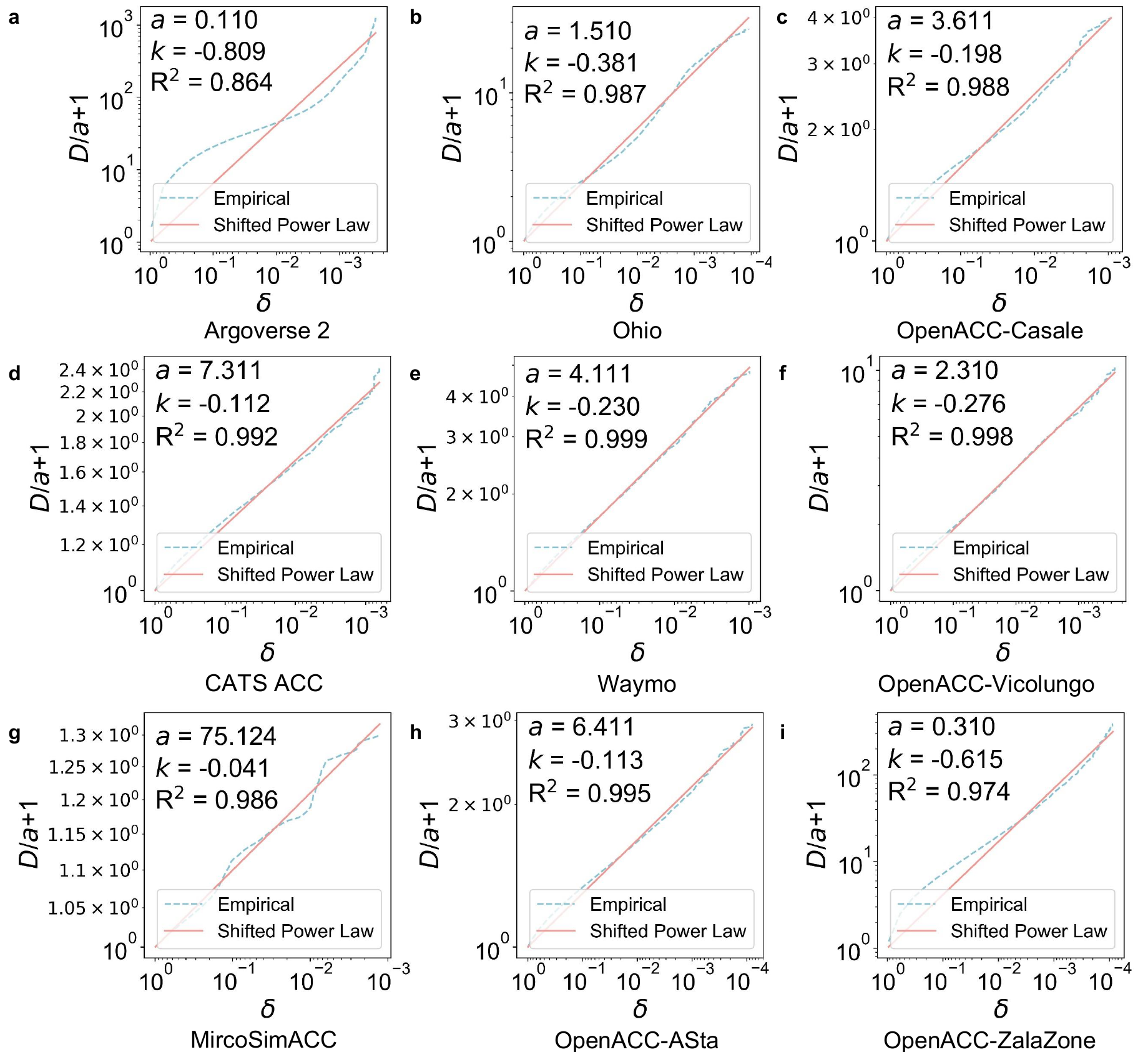}
\caption{\textbf{Fitting results of AVs across multiple datasets.}
\textbf{a–j}, AV longitudinal acceleration distributions from CATS ACC, Waymo, Argoverse~2, MicroSimACC, OpenACC (ASta, ZalaZone, Casale, Vicolungo), and Central Ohio datasets. 
The shifted power law (red) accurately reproduces empirical behavior (blue), achieving $\mathrm{R}^2>0.95$ across datasets, and demonstrating strong generalization across geographic regions and driving contexts.}
\label{fig:figC13}
\end{figure}


\newpage
\begin{figure}[!ht]
\centering
\includegraphics[width=1\linewidth]{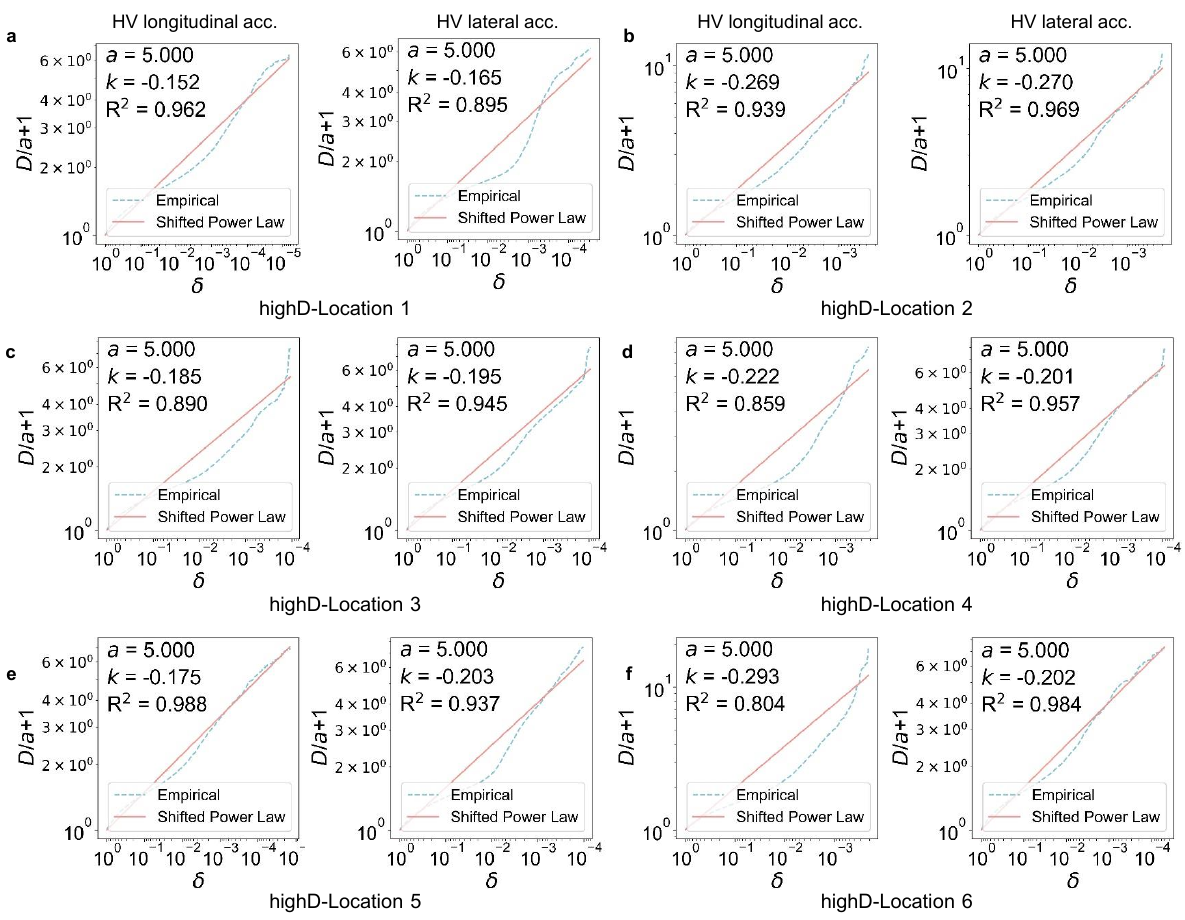}
\caption{\textbf{Fitting results of HVs on the highD dataset with fixed scale parameter $a=5$.}
\textbf{a–f}, HV longitudinal and lateral acceleration distributions across six highway locations in Germany.
Even with a fixed scale, the shifted power law (red) aligns closely with empirical data (blue), indicating that the exponent $k$ captures most of the behavioral variability across highway environments.}
\label{fig:figC21}
\end{figure}

\newpage
\begin{figure}[!ht]
\centering
\includegraphics[width=1\linewidth]{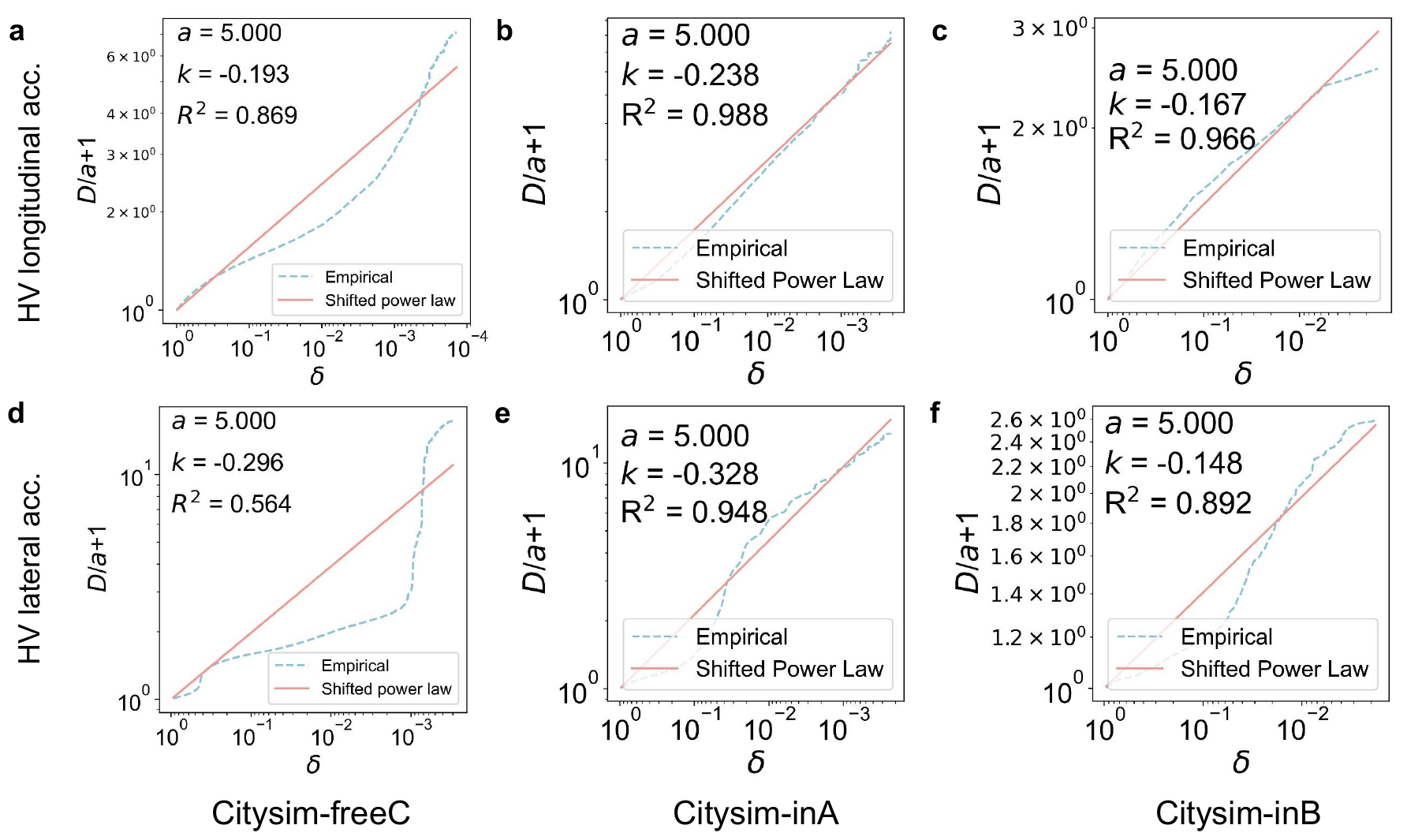}
\caption{\textbf{Fitting results of HVs on the CitySim dataset with fixed scale parameter $a=5$.}
\textbf{a–f}, HV longitudinal and lateral acceleration distributions from CitySim-freeC (China), CitySim-inA (non-signalized, U.S.), and CitySim-inB (signalized, U.S.) scenarios.
Despite fixing $a=5$, the shifted power law (red) maintains good agreement with empirical data (blue), reflecting consistent representation of stochastic urban driving behavior.}
\label{fig:figC22}
\end{figure}

\newpage
\begin{figure}[!ht]
\centering
\includegraphics[width=1\linewidth]{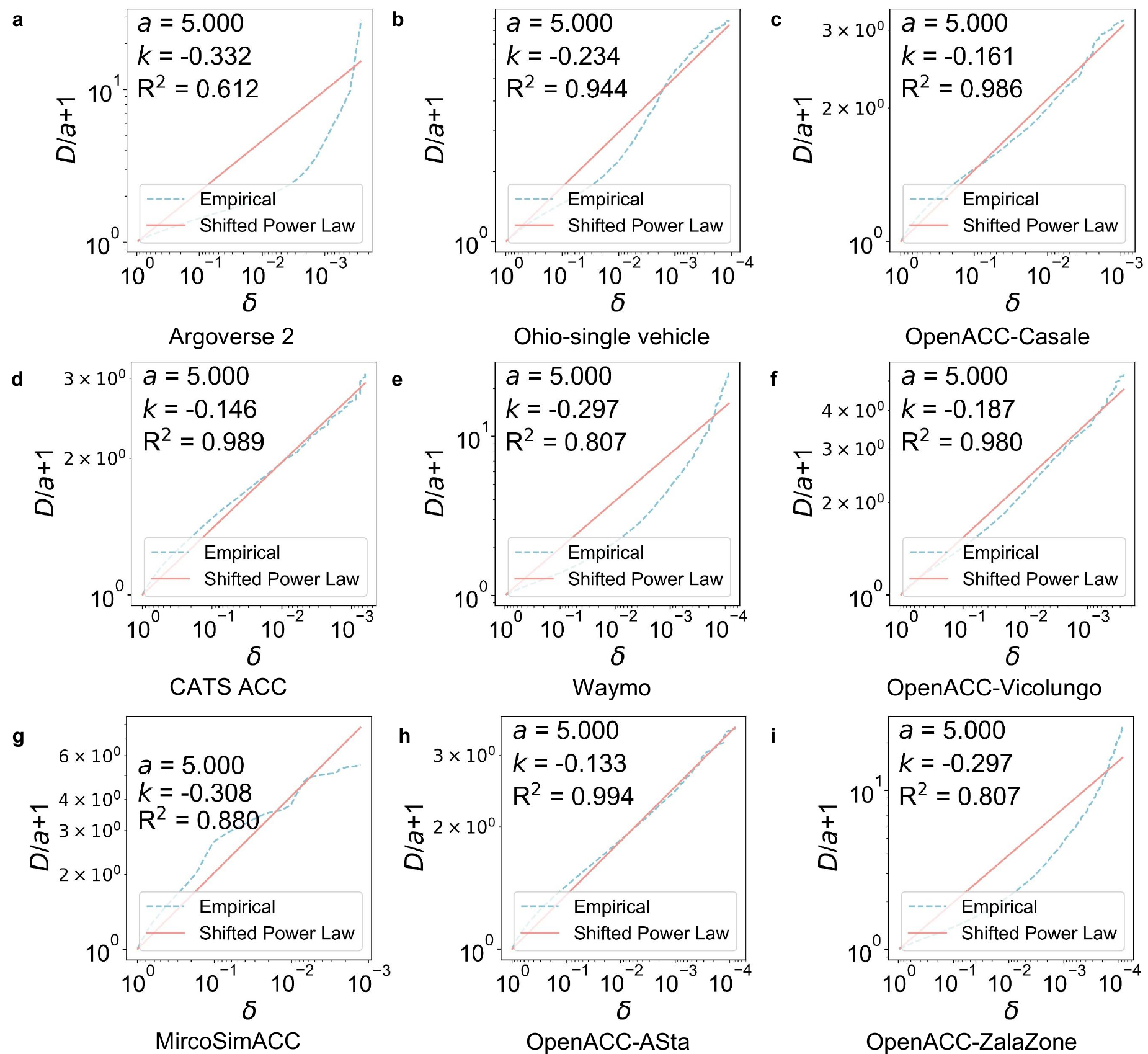}
\caption{\textbf{Fitting results of AVs across datasets with fixed scale parameter $a=5$.}
\textbf{a–j}, AV longitudinal acceleration distributions from CATS ACC, Waymo, Argoverse~2, MicroSimACC, OpenACC (ASta, ZalaZone, Casale, Vicolungo), and Central Ohio datasets.
The shifted power law (red) continues to reproduce the empirical patterns (blue) with high $\mathrm{R}^2$ values, demonstrating robustness and scalability under constrained parameterization.}
\label{fig:figC23}
\end{figure}


\newpage
\begin{figure}[!ht]
    \centering
    \includegraphics[width=1.0\linewidth]{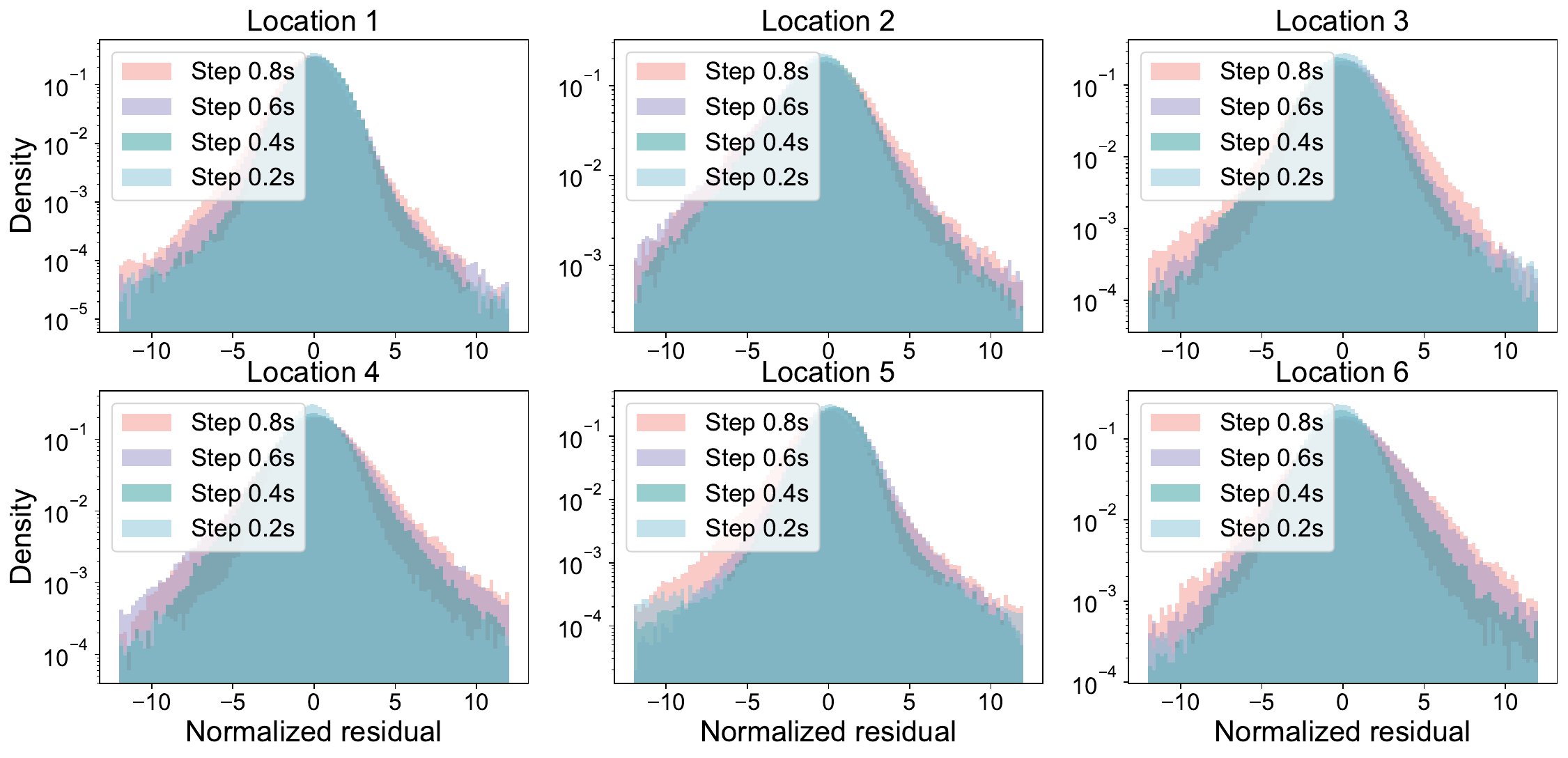}
    \caption{\textbf{Comparison of lateral normalized residual distributions of the highD dataset involving different prediction horizons, including 0.2\,s, 0.4\,s, 0.6\,s, and 0.8\,s.} To ensure a fair comparison, the input sequence length is adjusted to maintain a constant historical context (i.e., 2.4\,s) for all predictions. The results indicate that while predictions over longer horizons exhibit slightly greater dispersion, the overall shape and central tendency of the distributions remain consistent.}
    \label{figa: com_pred_steps_lat}
\end{figure}

\newpage
\begin{figure}[!ht]
    \centering
    \includegraphics[width=1.0\linewidth]{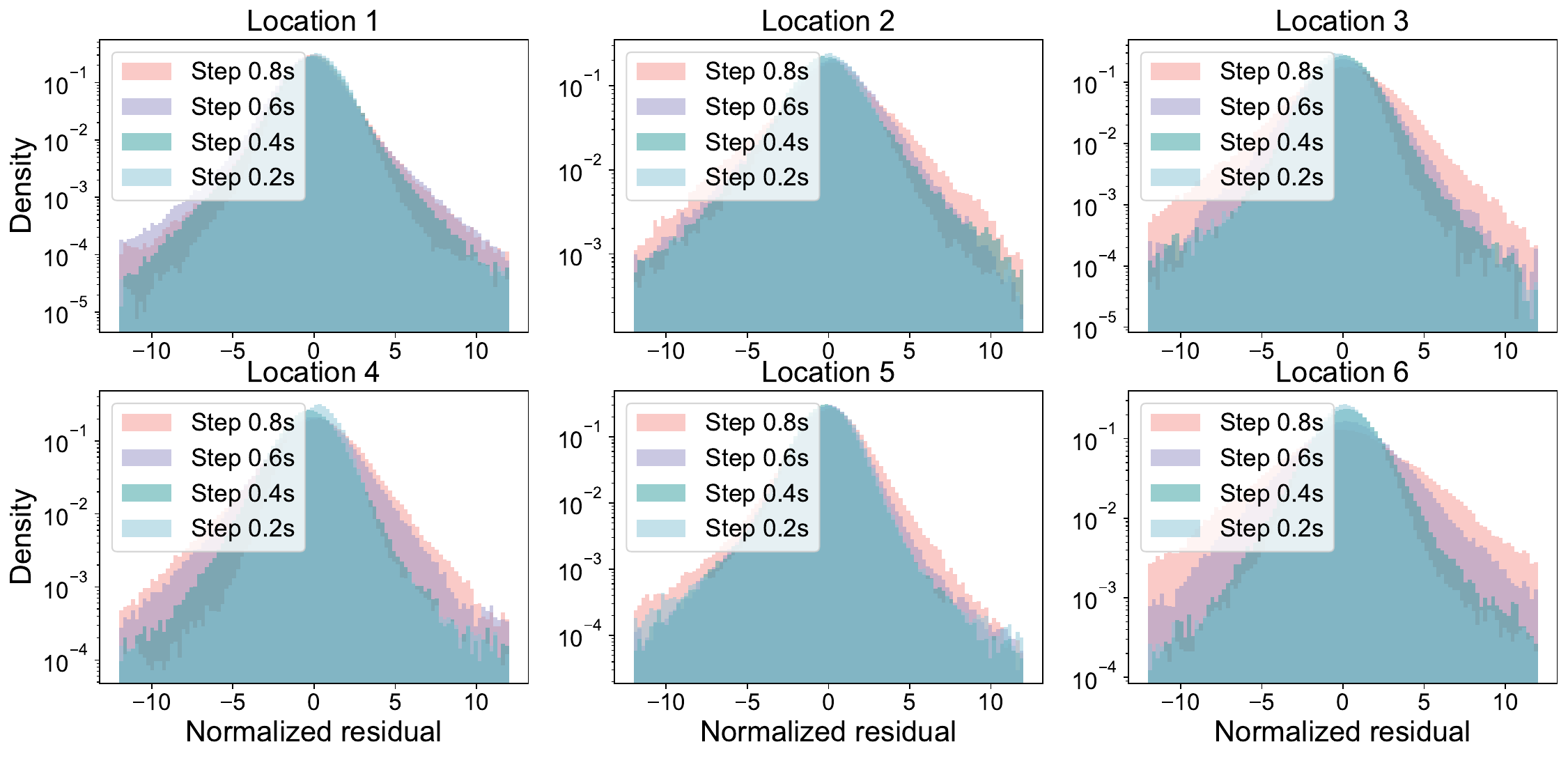}
    \caption{\textbf{Comparison of longitudinal normalized residual distributions of the highD dataset involving different prediction horizons, including 0.2\,s, 0.4\,s, 0.6\,s, and 0.8\,s.} To ensure a fair comparison, the input sequence length is adjusted to maintain a constant historical context (i.e., 2.4\,s) for all predictions. The results indicate that while predictions over longer horizons exhibit slightly greater dispersion, the overall shape and central tendency of the distributions remain consistent.}
    \label{figa: com_pred_steps_lon}
\end{figure}


\newpage
\begin{figure}[!ht]
    \centering
    \includegraphics[width=1.0\linewidth]{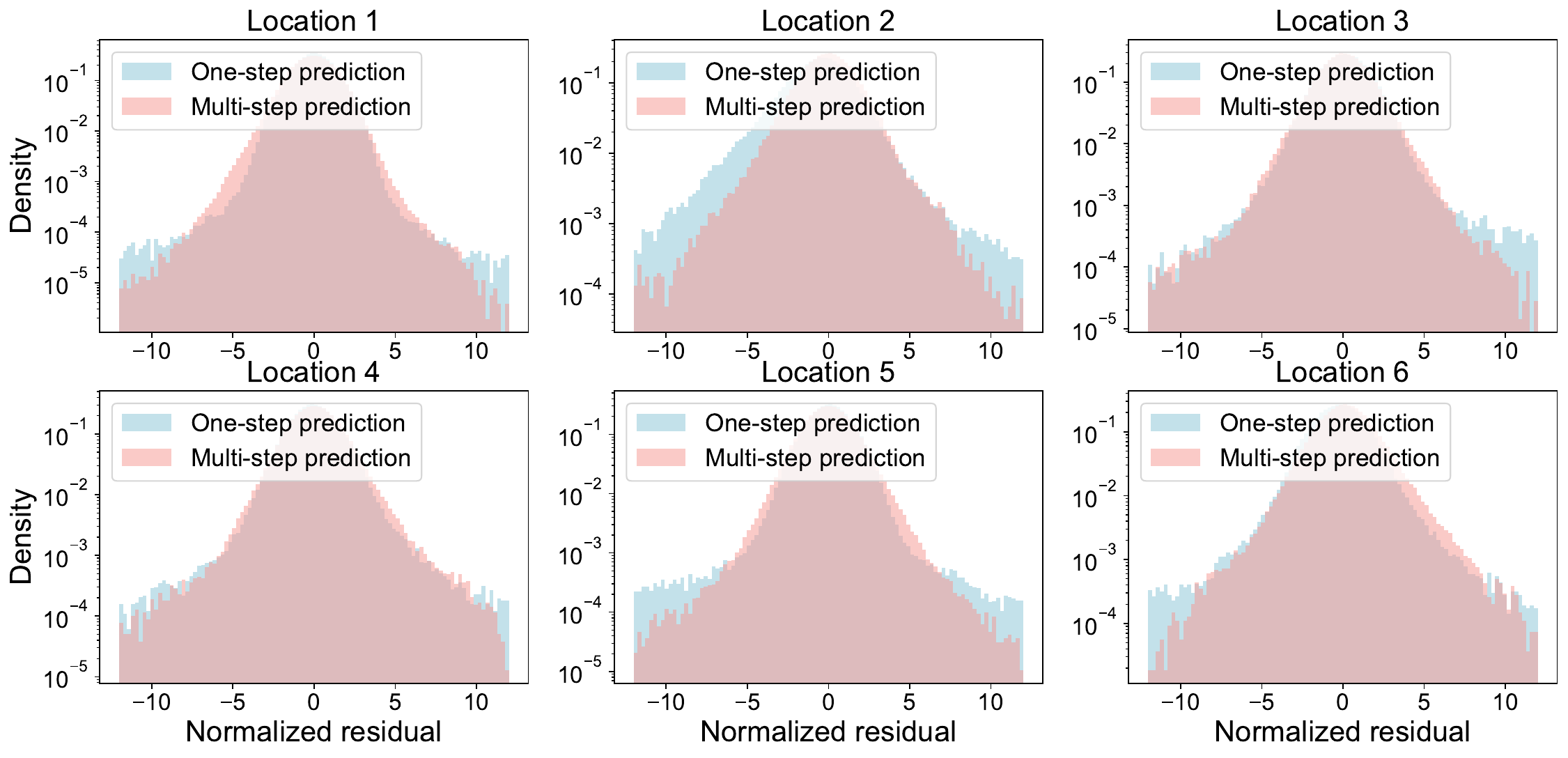}
    \caption{\textbf{Comparison of lateral normalized residual distributions of the highD dataset between one-step prediction and multi-step prediction at the next 0.2\,s time step.} The one-step prediction method is designed to forecast exclusively the immediate next time step, whereas the multi-step strategy is tasked with predicting a sequence of future states, including the target time step at 0.2~s.}
    \label{figa: com_pred_methods_2_lat}
\end{figure}

\newpage
\begin{figure}[!ht]
    \centering
    \includegraphics[width=1.0\linewidth]{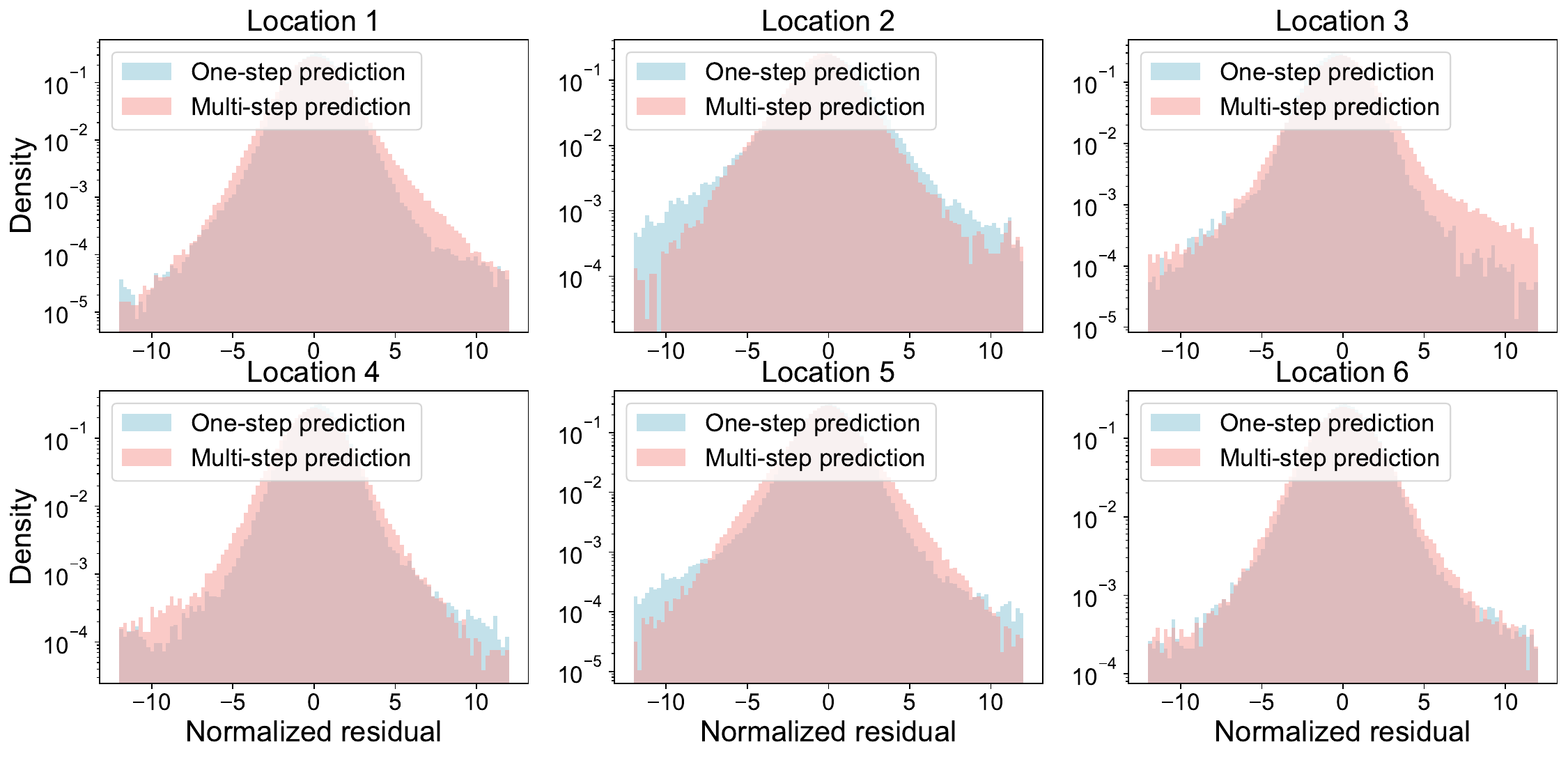}
    \caption{\textbf{Comparison of longitudinal normalized residual distributions of the highD dataset between one-step prediction and multi-step prediction at the next 0.2\,s time step.} The one-step prediction method is designed to forecast exclusively the immediate next time step, whereas the multi-step strategy is tasked with predicting a sequence of future states, including the target time step at 0.2~s.}
    \label{figa: com_pred_methods_2_lon}
\end{figure}

\newpage
\begin{figure}[!ht]
    \centering
    \includegraphics[width=1.0\linewidth]{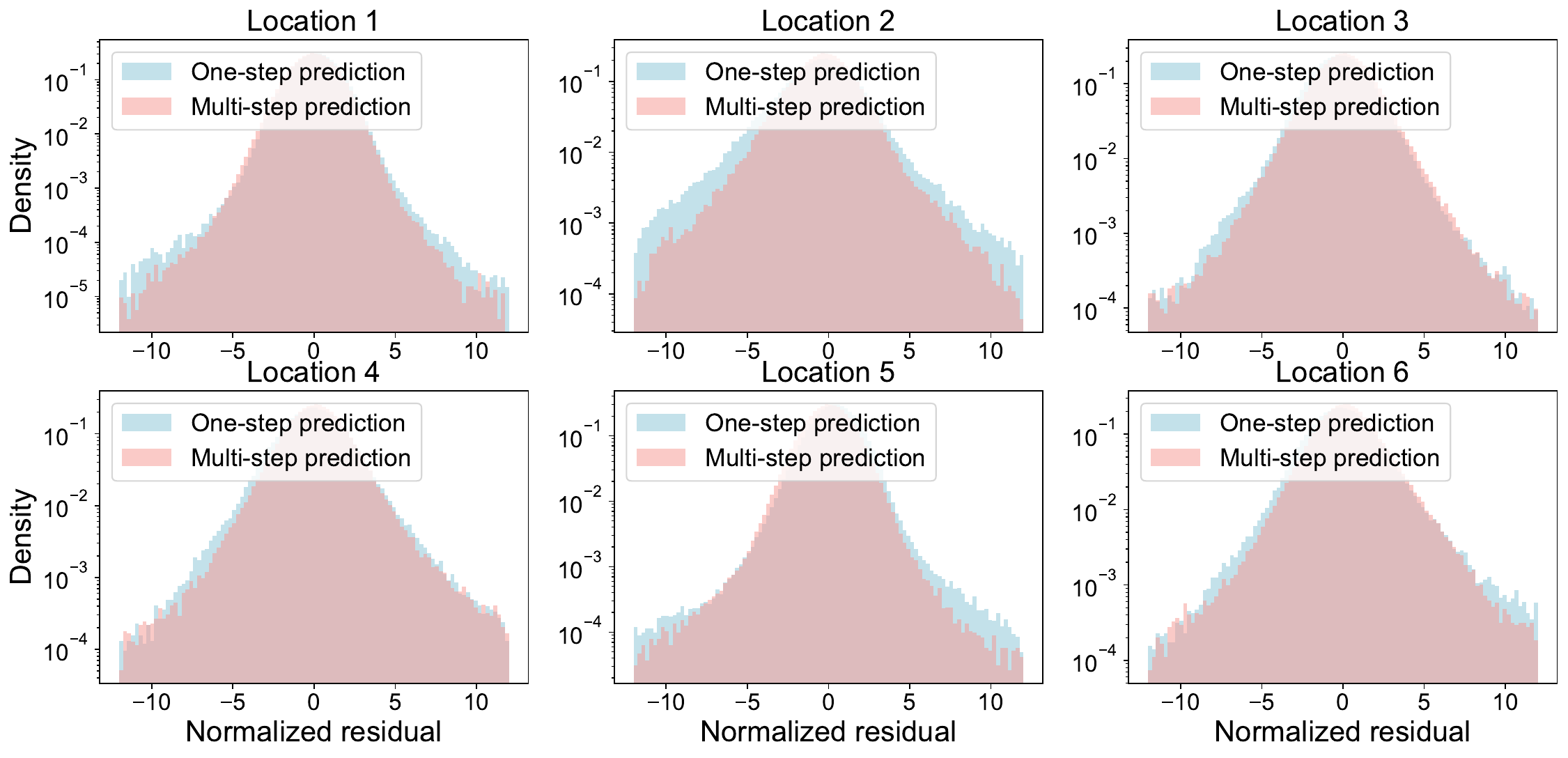}
    \caption{\textbf{Comparison of lateral normalized residual distributions of the highD dataset between one-step prediction and multi-step prediction at the next 0.4\,s time step.} The one-step prediction method is designed to forecast exclusively the immediate next time step, whereas the multi-step strategy is tasked with predicting a sequence of future states, including the target time step at 0.4~s.}
    \label{figa: com_pred_methods_4_lat}
\end{figure}

\newpage
\begin{figure}[!ht]
    \centering
    \includegraphics[width=1.0\linewidth]{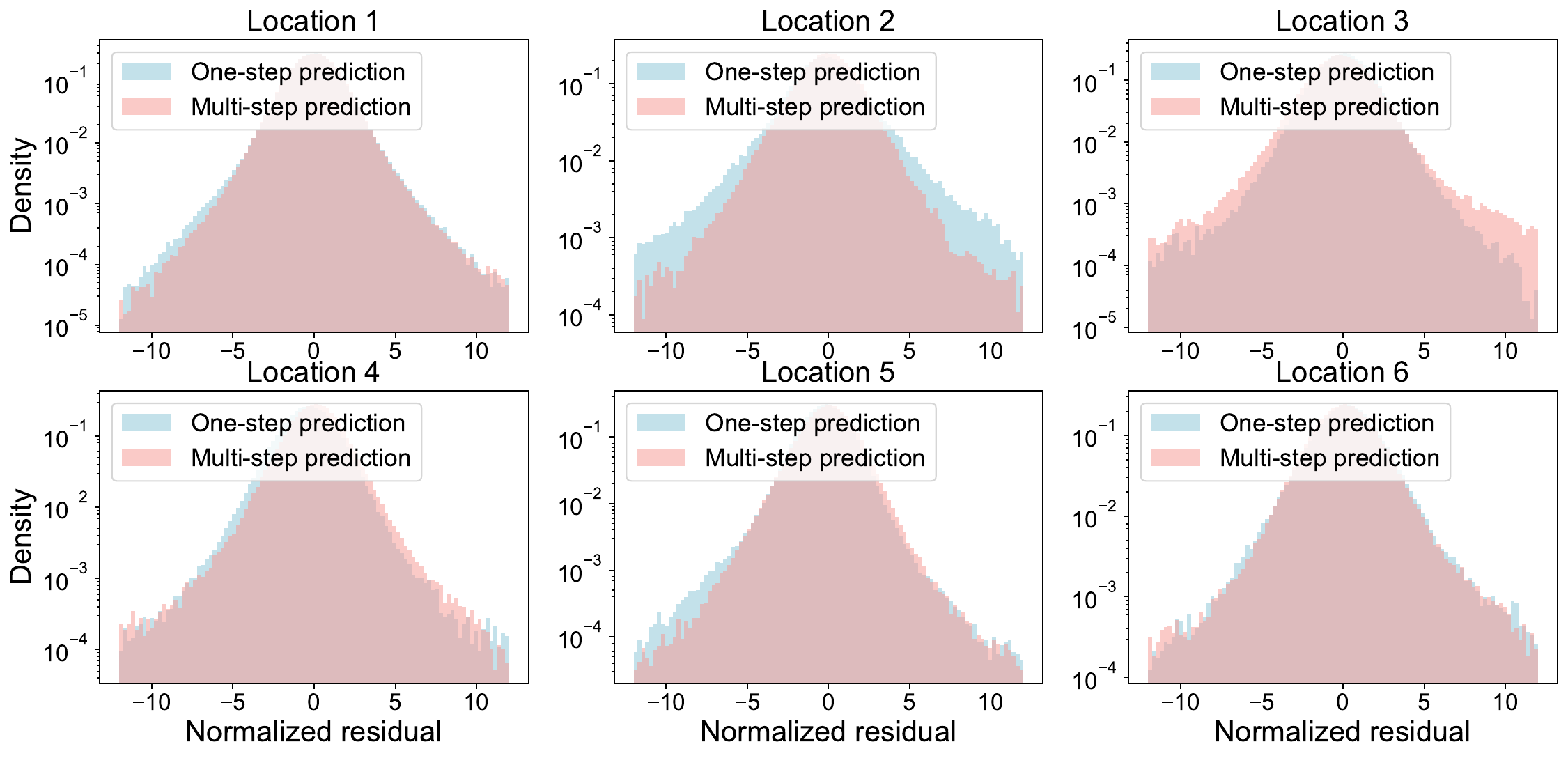}
    \caption{\textbf{Comparison of longitudinal normalized residual distributions of the highD dataset between one-step prediction and multi-step prediction at the next 0.4\,s time step.} The one-step prediction method is designed to forecast exclusively the immediate next time step, whereas the multi-step strategy is tasked with predicting a sequence of future states, including the target time step at 0.4~s.}
    \label{figa: com_pred_methods_4_lon}
\end{figure}

\newpage
\begin{figure}[!ht]
    \centering
    \includegraphics[width=1.0\linewidth]{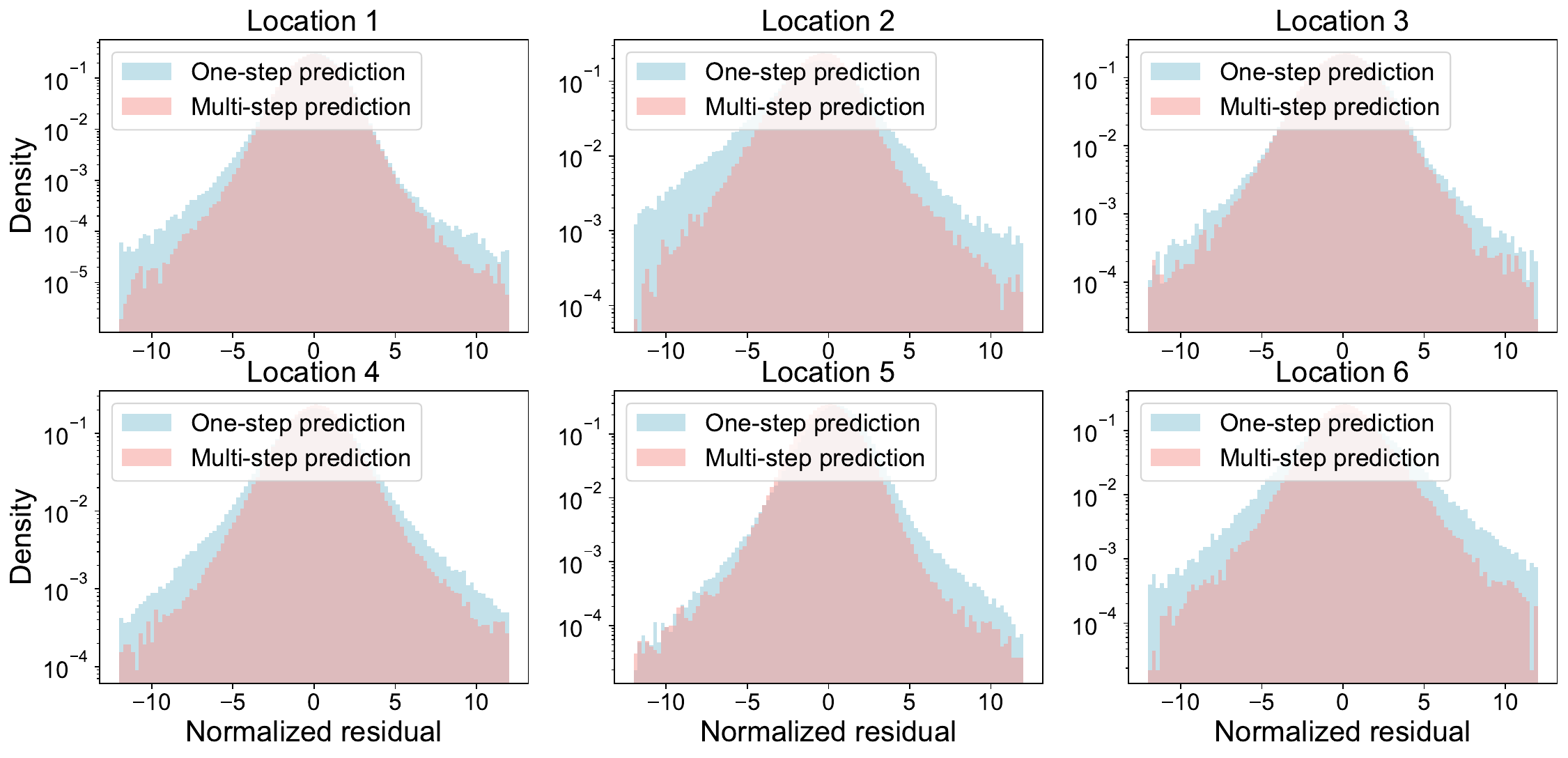}
    \caption{\textbf{Comparison of lateral normalized residual distributions of the highD dataset between one-step prediction and multi-step prediction at the next 0.6\,s time step.} The one-step prediction method is designed to forecast exclusively the immediate next time step, whereas the multi-step strategy is tasked with predicting a sequence of future states, including the target time step at 0.6~s.}
    \label{figa: com_pred_methods_6_lat}
\end{figure}

\newpage
\begin{figure}[!ht]
    \centering
    \includegraphics[width=1.0\linewidth]{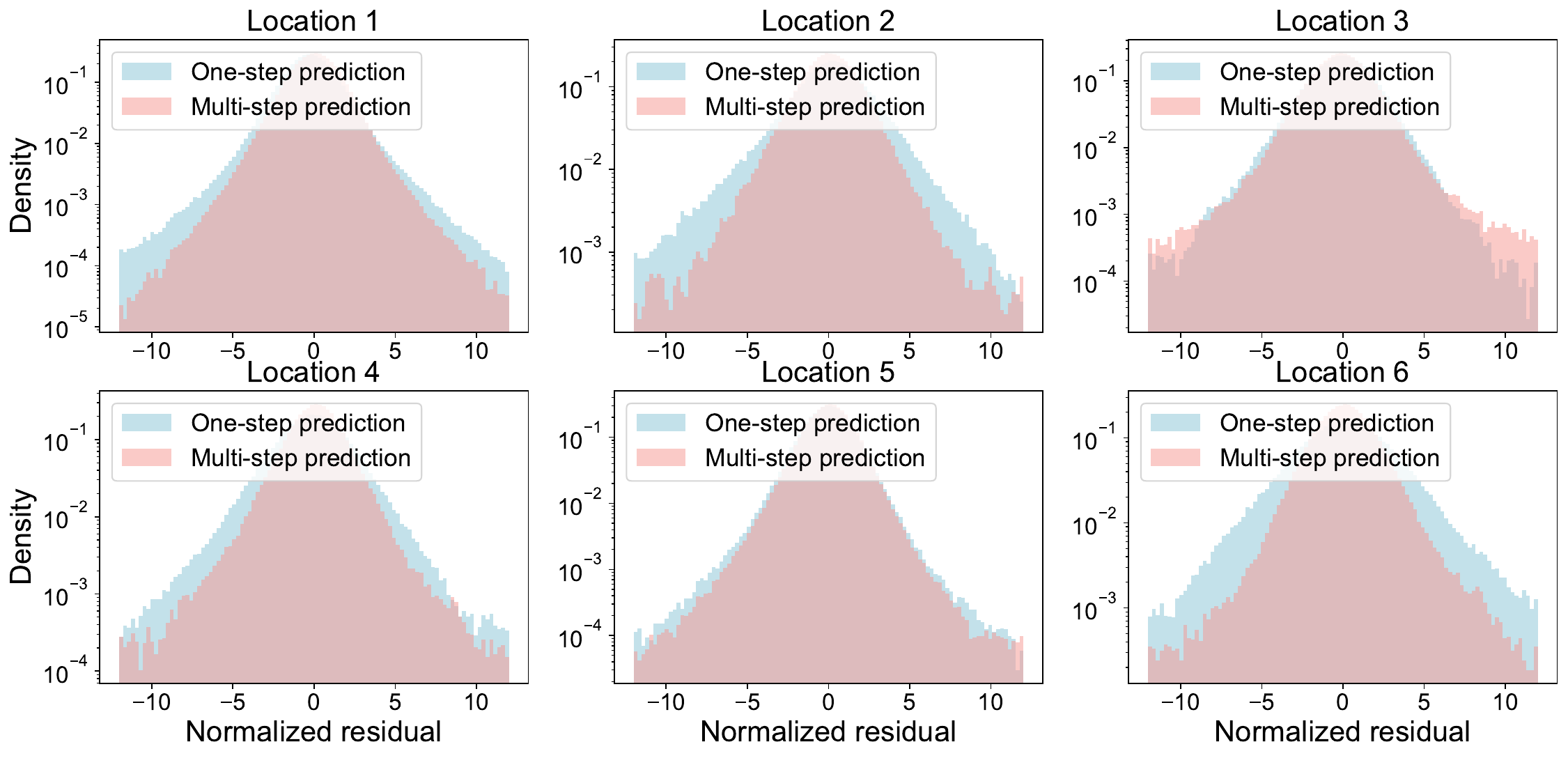}
    \caption{\textbf{Comparison of longitudinal normalized residual distributions of the highD dataset between one-step prediction and multi-step prediction at the next 0.6\,s time step.} The one-step prediction method is designed to forecast exclusively the immediate next time step, whereas the multi-step strategy is tasked with predicting a sequence of future states, including the target time step at 0.6~s.}
    \label{figa: com_pred_methods_6_lon}
\end{figure}

\newpage
\begin{figure}[!ht]
    \centering
    \includegraphics[width=1.0\linewidth]{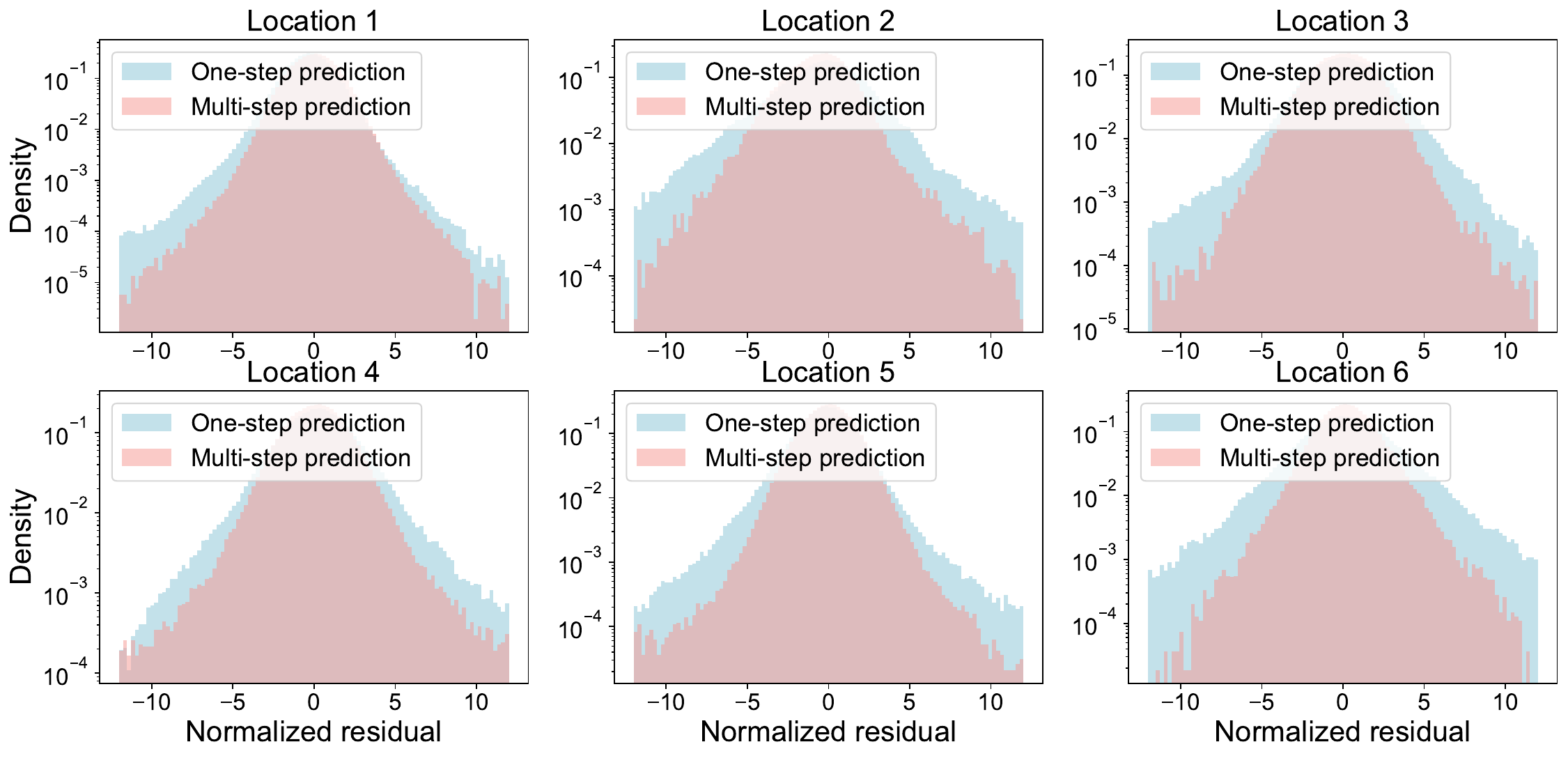}
    \caption{\textbf{Comparison of lateral normalized residual distributions of the highD dataset between one-step prediction and multi-step prediction at the next 0.8\,s time step.} The one-step prediction method is designed to forecast exclusively the immediate next time step, whereas the multi-step strategy is tasked with predicting a sequence of future states, including the target time step at 0.8~s.}
    \label{figa: com_pred_methods_8_lat}
\end{figure}

\newpage
\begin{figure}[!ht]
    \centering
    \includegraphics[width=1.0\linewidth]{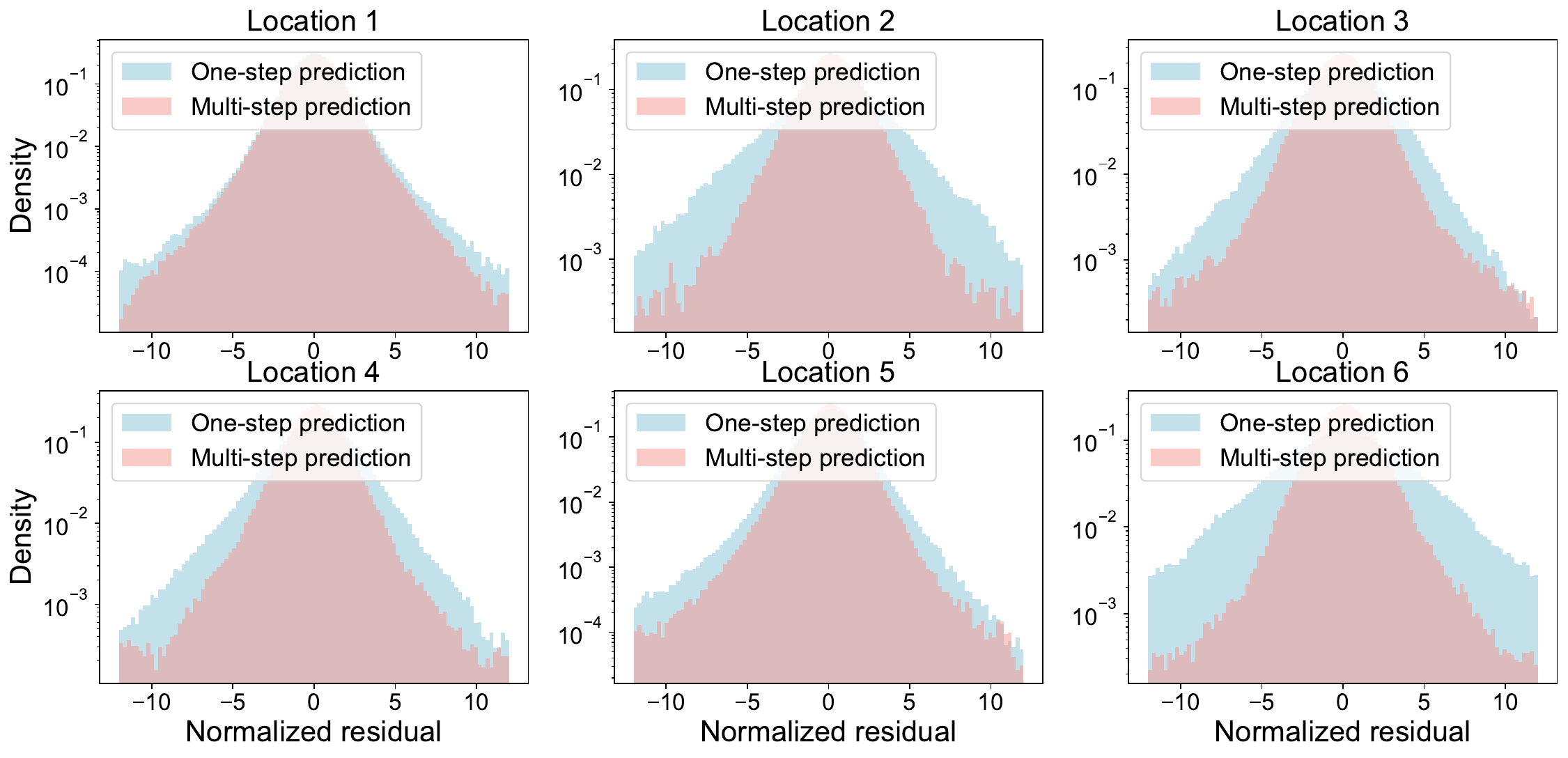}
    \caption{\textbf{Comparison of longitudinal normalized residual distributions of the highD dataset between one-step prediction and multi-step prediction at the next 0.8\,s time step.} The one-step prediction method is designed to forecast exclusively the immediate next time step, whereas the multi-step strategy is tasked with predicting a sequence of future states, including the target time step at 0.8~s.}
    \label{figa: com_pred_methods_8_lon}
\end{figure}


\newpage
\section{Supplementary tables}


\begin{table}[!ht]
\centering
\caption{Measurement results of the lateral direction of the highD dataset using different metrics.}
\label{taba: metrics_highD_lat}
\begin{tabular}{c l r r r}
\toprule
Location & Method & RP5 & Log-Likelihood & KL Divergence \\
\midrule
Location 1 & Gaussian & 5315.679 & -466.459 & 0.057 \\
\rowcolor{lightgray!30}
Location 1 & Shifted Power Law & 0.281 & -249.116 & 0.294 \\
Location 2 & Gaussian & 185623.174 & -466.459 & 1.054 \\
\rowcolor{lightgray!30}
Location 2 & Shifted Power Law & 1.394 & -196.227 & 0.172 \\
Location 3 & Gaussian & 30086.391 & -466.459 & 0.242 \\
\rowcolor{lightgray!30}
Location 3 & Shifted Power Law & 0.661 & -223.996 & 0.211 \\
Location 4 & Gaussian & 30876.381 & -466.459 & 0.201 \\
\rowcolor{lightgray!30}
Location 4 & Shifted Power Law & 0.604 & -221.018 & 0.156 \\
Location 5 & Gaussian & 20255.381 & -466.459 & 0.129 \\
\rowcolor{lightgray!30}
Location 5 & Shifted Power Law & 0.451 & -225.581 & 0.194 \\
Location 6 & Gaussian & 52832.955 & -466.459 & 0.382 \\
\rowcolor{lightgray!30}
Location 6 & Shifted Power Law & 0.630 & -206.338 & 0.112 \\
\midrule
Average & Gaussian &  54164.994 & -466.459 & 0.344 \\
\rowcolor{lightgray!30}
Average & Shifted Power Law &  0.670 & -220.379 & 0.190 \\
\bottomrule
\end{tabular}
\end{table}

\newpage
\begin{table}[!ht]
\centering
\caption{Measurement results of the longitudinal direction of the highD dataset using different metrics.}
\label{taba: metrics_highD_lon}
\begin{tabular}{c l r r r}
\toprule
Location & Method & RP5 & Log-Likelihood & KL Divergence \\
\midrule
Location 1 & Gaussian & 12545.302 & -466.459 & 0.125 \\
\rowcolor{lightgray!30}
Location 1 & Shifted Power Law & 0.628 & -244.121 & 0.226 \\
Location 2 & Gaussian & 122777.672 & -466.459 & 0.733 \\
\rowcolor{lightgray!30}
Location 2 & Shifted Power Law & 1.259 & -205.360 & 0.193 \\
Location 3 & Gaussian & 21149.280 & -466.459 & 0.239 \\
\rowcolor{lightgray!30}
Location 3 & Shifted Power Law & 0.784 & -239.682 & 0.387 \\
Location 4 & Gaussian & 23214.900 & -466.459 & 0.170 \\
\rowcolor{lightgray!30}
Location 4 & Shifted Power Law & 0.654 & -234.926 & 0.379 \\
Location 5 & Gaussian & 31602.958 & -466.459 & 0.216 \\
\rowcolor{lightgray!30}
Location 5 & Shifted Power Law & 0.578 & -216.511 & 0.103 \\
Location 6 & Gaussian & 52757.621 & -466.459 & 0.388 \\
\rowcolor{lightgray!30}
Location 6 & Shifted Power Law & 1.184 & -229.880 & 0.462 \\
\midrule
Average & Gaussian &  44007.956 & -466.459 & 0.312 \\
\rowcolor{lightgray!30}
Average & Shifted Power Law &  0.848 & -228.413 & 0.292 \\
\bottomrule
\end{tabular}
\end{table}

\newpage
\begin{table}[!ht]
\centering
\caption{Measurement results of the CitySim dataset in the lateral direction (including Intersection A, Intersection B, and Freeway C).}
\label{taba: metrics_CitySim_lat}
\begin{tabular}{l l r r r}
\toprule
Location & Method & RP5 & Log-Likelihood & KL Divergence \\
\midrule
Intersection A & Gaussian & 78300.474 & -466.459 & 0.401 \\
\rowcolor{lightgray!30}
Intersection A & Shifted Power Law & 0.167 & -165.115 & 0.525 \\
Intersection B & Gaussian & 54974.100 & -466.459 & 0.538 \\
\rowcolor{lightgray!30}
Intersection B & Shifted Power Law & 1.863 & -240.429 & 0.042 \\
Freeway C & Gaussian & 40690.395 & -466.459 & 0.872 \\
\rowcolor{lightgray!30}
Freeway C & Shifted Power Law & 1.145 & -236.552 & 0.518 \\
\midrule
Average & Gaussian &  57988.323 & -466.459 & 0.604 \\
\rowcolor{lightgray!30}
Average & Shifted Power Law &  1.058 & -214.032 & 0.362 \\
\bottomrule
\end{tabular}
\end{table}

\newpage
\begin{table}[!ht]
\centering
\caption{Measurement results of the CitySim dataset in the longitudinal direction (including Intersection A, Intersection B, and Freeway C).}
\label{taba: metrics_CitySim_lon}
\begin{tabular}{l l r r r}
\toprule
Location & Method & RP5 & Log-Likelihood & KL Divergence \\
\midrule
Intersection A & Gaussian & 121405.953 & -466.459 & 0.525 \\
\rowcolor{lightgray!30}
Intersection A & Shifted Power Law & 0.940 & -196.108 & 0.049 \\
Intersection B & Gaussian & 84948.236 & -466.459 & 0.504 \\
\rowcolor{lightgray!30}
Intersection B & Shifted Power Law & 1.056 & -202.221 & 0.031 \\
Freeway C & Gaussian & 21648.627 & -466.459 & 0.127 \\
\rowcolor{lightgray!30}
Freeway C & Shifted Power Law & 0.781 & -240.430 & 0.318 \\
\midrule
Average & Gaussian &  76000.939 & -466.459 & 0.385 \\
\rowcolor{lightgray!30}
Average & Shifted Power Law &  0.926 & -212.920 & 0.133 \\
\bottomrule
\end{tabular}
\end{table}

\newpage

\begin{table}[!ht]
\centering
\caption{Measurement results of the AV datasets using different metrics.}
\label{taba: metrics_AV}
\begin{tabular}{l l r r r}
\toprule
Dataset & Method & RP5 & Log-Likelihood & KL Divergence \\
\midrule
Argoverse~2 & Gaussian & 30096.108 & -466.459 & 0.159 \\
\rowcolor{lightgray!30}
Argoverse & Shifted Power Law & 1.479 & -246.771 & 0.541 \\
CATSACC & Gaussian & 34245.690 & -466.459 & 0.235 \\
\rowcolor{lightgray!30}
CATSACC & Shifted Power Law & 0.822 & -220.463 & 0.040 \\
MicroSimACC & Gaussian & 526545.139 & -466.459 & 2.716 \\
\rowcolor{lightgray!30}
MicroSimACC & Shifted Power Law & 0.783 & -157.886 & 0.087 \\
Ohio (single vehicle) & Gaussian & 48225.987 & -466.459 & 0.222 \\
\rowcolor{lightgray!30}
Ohio (single vehicle) & Shifted Power Law & 0.673 & -213.204 & 0.066 \\
Ohio (two vehicle) & Gaussian & 74935.487 & -466.459 & 0.314 \\
\rowcolor{lightgray!30}
Ohio (two vehicle) & Shifted Power Law & 0.660 & -200.154 & 0.008 \\
OA-ASta & Gaussian & 22889.371 & -466.459 & 0.145 \\
\rowcolor{lightgray!30}
OA-ASta & Shifted Power Law & 0.857 & -231.733 & 0.057 \\
OA-Casale & Gaussian & 37107.737 & -466.459 & 0.204 \\
\rowcolor{lightgray!30}
OA-Casale & Shifted Power Law & 0.748 & -218.676 & 0.074 \\
OA-Vicolungo & Gaussian & 54715.798 & -466.459 & 0.224 \\
\rowcolor{lightgray!30}
OA-Vicolungo & Shifted Power Law & 0.950 & -216.876 & 0.038 \\
OA-ZalaZone & Gaussian & 46300.897 & -466.459 & 0.185 \\
\rowcolor{lightgray!30}
OA-ZalaZone & Shifted Power Law & 1.703 & -243.726 & 0.260 \\
Waymo & Gaussian & 115573.915 & -466.459 & 0.538 \\
\rowcolor{lightgray!30}
Waymo & Shifted Power Law & 0.984 & -197.339 & 0.010 \\
\midrule
Average & Gaussian &  99063.613 & -466.459 & 0.494 \\
Average & Shifted Power Law &  0.966 & -214.683 & 0.118 \\
\bottomrule
\end{tabular}
\end{table}

\newpage
\begin{table}[!ht]
\centering
\caption{Measurement results of the highD dataset among different long-tail distributions. ``Lat'' and ``Lon'' represent lateral and longitudinal directions, respectively, and ``df'' denotes the degree of freedom.}
\label{taba: metrics_others_highD}
\begin{tabular}{l l r r r}
\toprule
Method & Direction & RP5 & Log-Likelihood & KL Divergence \\
\midrule
Laplace & Lat & 14.509 & -276.256 & 0.209 \\
Student's t (df=3) & Lat & 4.345 & -257.546 & 0.322 \\
Student's t (df=4) & Lat & 6.131 & -260.988 & 0.231 \\
\rowcolor{lightgray!30}
Shifted Power Law & Lat & 0.670 & -220.379 & 0.190 \\
\midrule
Laplace & Lon & 11.788 & -276.256 & 0.203 \\
Student's t (df=3) & Lon & 3.530 & -257.546 & 0.323 \\
Student's t (df=4) & Lon & 4.982 & -260.988 & 0.228 \\
\rowcolor{lightgray!30}
Shifted Power Law & Lon & 0.848 & -228.413 & 0.292 \\
\bottomrule
\end{tabular}
\end{table}


\newpage
\begin{table}[!ht]
\centering
\caption{Measurement results of the CitySim dataset among different long-tail distributions. ``Lat'' and ``Lon'' represent lateral and longitudinal directions, respectively, and ``df'' denotes the degree of freedom.}
\label{taba: metrics_others_CitySim}
\begin{tabular}{l l r r r}
\toprule
Method & Direction & RP5 & Log-Likelihood & KL Divergence \\
\midrule
Laplace & Lat & 15.532 & -276.256 & 0.271 \\
Student's t (df=3) & Lat & 4.652 & -257.546 & 0.280 \\
Student's t (df=4) & Lat & 6.564 & -260.988 & 0.312 \\
\rowcolor{lightgray!30}
Shifted Power Law & Lat & 1.058 & -214.032 & 0.362 \\
\midrule
Laplace & Lon & 20.358 & -276.256 & 0.136 \\
Student's t (df=3) & Lon & 6.097 & -257.546 & 0.190 \\
Student's t (df=4) & Lon & 8.603 & -260.988 & 0.148 \\
\rowcolor{lightgray!30}
Shifted Power Law & Lon & 0.926 & -212.920 & 0.133 \\
\bottomrule

\end{tabular}
\end{table}

\newpage
\begin{table}[!ht]
\centering
\caption{Measurement results of the AV datasets among different long-tail distributions. ``df'' denotes the degree of freedom.}
\label{taba: metrics_others_AV}
\begin{tabular}{l l r r r}
\toprule
Method & RP5 & Log-Likelihood & KL Divergence \\
\midrule
Laplace & 26.535 & -276.256 & 0.201 \\
Student's t (df=3) & 7.946 & -257.546 & 0.262 \\
Student's t (df=4) & 11.214 & -260.988 & 0.206 \\
\rowcolor{lightgray!30}
Shifted Power Law & 0.966 & -214.683 & 0.118 \\
\bottomrule
\end{tabular}
\end{table}



\newpage
\section{Details of datasets}
\label{sec:dataset_details}

To facilitate reproducibility and comparative studies, this section provides detailed descriptions of all datasets used in this study.
\begin{itemize}
    \item highD (Germany, HV) \cite{krajewski2018highd}: The highD dataset comprises naturalistic vehicle trajectories recorded on German highways using aerial drones. Spanning 16.5 hours of highway data across six locations, it captures detailed vehicle interactions, including car-following and lane-changing behaviors. The dataset encompasses over 110,000 vehicles and provides high-resolution data with a sampling rate of 25 Hz, offering precise measurements of vehicle positions, velocities, and accelerations.
    \item CitySim (USA, China, HV) \cite{zheng2024citysim}: CitySim is a drone-recorded vehicle trajectory dataset designed for safety research and digital twin applications. It contains over 1,140 minutes of annotated trajectories across 12 urban scenarios, including freeways, weaving zones, and both signalized and unsignalized intersections. The data were captured using DJI drones at heights of 120–320 meters and recorded at 30 FPS in up to 5K resolution. The dataset provides frame-level bounding boxes, headings, velocities, and includes over 8,800 cut-ins and numerous merge and diverge events. A multi-stage pipeline involving stabilization, detection, tracking, and manual correction yields high-accuracy trajectories with a post-processed IoU of 0.976. CitySim also offers surrogate safety metrics and geo-aligned 3D maps with signal data, supporting applications in rare-event modeling and AV safety validation.
    \item MicroSimACC Dataset (USA, AV) \cite{yang2024microsimacc}: The MicroSimACC dataset was collected using two to three Toyota Corolla vehicles equipped with Adaptive Cruise Control (ACC) at Florida Atlantic University. Vehicle trajectories were recorded through OBD-II loggers at 5 Hz, providing accurate measurements of speed, acceleration, and distance traveled. Field experiments were conducted on isolated public roads in Delray Beach and Loxahatchee, Florida, covering a wide range of traffic conditions with a focus on a common vehicle model under controlled settings.
    \item CATS Dataset (USA, AV) \cite{shi2021empirical,ma2025real}: The CATS dataset contains car-following trajectories of vehicle platoons with ACC engaged. It includes two parts: the CATS ACC dataset collected on urban roads in Tampa, Florida, and the CATS UWM dataset collected on suburban roads in Madison, Wisconsin. The dataset provides detailed position and speed measurements for analyzing platoon dynamics under real-world driving conditions.
    \item OpenACC Dataset (Europe, AV) \cite{makridis2021openacc}:     Collected by the European Commission, the OpenACC dataset documents the behavior of commercial ACC systems under various driving conditions. It consists of four sub-datasets:
    \begin{itemize}
        \item Casale Dataset (Italy, 2018): Two- and three-vehicle platoons tested on Italian freeways, with leaders introducing small perturbations around free-flow speed.
        \item Vicolungo Dataset (Italy, 2019): Three days of car-following tests involving five commercial vehicles of different makes and models, with leaders driving manually and followers using ACC when possible.
        \item AstaZero Dataset (Sweden, 2019): Controlled car-following tests on a 3.5-mile rural test track, with platoons driving at constant speeds or under target-speed perturbations, recorded by an inertial navigation system.
        \item ZalaZONE Dataset (Hungary, 2019): Car-following experiments with seven vehicles on circular and handling tracks, including perturbation tests and intersection stopping behavior with different ACC gap settings.
    \end{itemize}
    \item Central Ohio Dataset (USA, AV) \cite{xia2023automated}: The Central Ohio dataset captures naturalistic driving behavior of Level 2 ADAS-equipped vehicles on urban and suburban roadways in Ohio. Two types of vehicles were used: readily identifiable vehicles with visible sensor stacks and discreet vehicles with concealed sensors. Both were equipped with cameras and LiDAR to record trajectories of surrounding traffic, enabling detailed analysis of adjacent vehicle interactions. This dataset includes two types of experiments: single-vehicle and two-vehicle settings. The main text presents results based on the single-vehicle dataset, while the appendix provides results for both settings.
    \item Waymo Open Dataset (USA, AV) \cite{sun2020scalability,hu2022processing}: The Waymo Open Dataset provides high-resolution multimodal sensor data collected by Waymo’s AVs across diverse environments, traffic conditions, and weather scenarios. Data were gathered in multiple U.S. cities, including San Francisco, Mountain View, Los Angeles, Phoenix, Detroit, and Seattle, offering a broad representation of urban and suburban driving contexts.
    \item Argoverse~2 Dataset (USA, AV) \cite{wilson2023argoverse}: The Argoverse~2 dataset provides large-scale trajectory data collected from multiple U.S. cities, including Miami, Pittsburgh, Austin, and Washington, D.C. It contains over 250,000 scenarios with detailed annotations of vehicle and pedestrian trajectories, map features, and contextual information. Designed for trajectory prediction tasks, the dataset supports research on long-horizon motion forecasting in diverse urban environments with heterogeneous traffic participants.
\end{itemize}

\newpage
\section{Prediction details}\label{seca: prediction}

In this section, we introduce the details of predicting the acceleration of the ego vehicle. As illustrated in Figure~\ref{figa: prediction_state}, the input at each time step is a fixed-length feature vector comprising the states of the ego vehicle and its surrounding vehicles. For the ego vehicle, we include its longitudinal and lateral velocities and accelerations, together with lane identifiers. For each of the eight neighboring vehicles, we extract a five-dimensional feature vector comprising longitudinal and lateral velocities, longitudinal and lateral accelerations, and the relative position with respect to the ego vehicle (as shown in Figure~\ref{figa: prediction_state}). These per-vehicle features are concatenated with the ego vehicle's feature into a fixed-length vector, yielding the feature dimension $F$ at each time step. Over a temporal observation window of $T$ steps, the resulting sequence is represented as an input tensor of shape $(B, T, F)$, where $B$ denotes the batch size.

As shown in Figure~\ref{figa: predict_network}, the prediction network consists of an LSTM encoder and a dual-head regression module. We use an LSTM with 2 layers and 128 hidden units per layer; the last hidden state is fed to two feed-forward networks (FFNs) that output the next-step mean $\hat{y}_{T+1}$ and the standard deviation $\hat{\gamma}_{T+1}$. Building on this, the model is trained under a one-step prediction scheme: given the observation window $x_{1:T}$, the network predicts only the immediate next acceleration at $T+1$. During inference, the predicted value can be recursively fed back as input to roll forward over longer horizons.

\newpage
\section{More results}

In this section, we report all the fitting results in \ref{seca: fitting_results} and those with a fixed scale ($a=5$) in \ref{seca: a5_results}.

\subsection{Fitting results}\label{seca: fitting_results}

As shown in Figures~\ref{fig:figC11}–\ref{fig:figC13}, the shifted power law provides robust fits to both HV and AV driving behaviors across multiple datasets. Figure~\ref{fig:figC11} presents the highD highway driving data collected in Germany, where subplots \textbf{a–f} depict HV longitudinal and lateral acceleration distributions across six recording locations. The fitted parameters $(a, k)$ and $\mathrm{R}^2$ show high tail fidelity, confirming the model’s ability to capture consistent stochastic patterns across distinct high-speed highway environments.
Figure~\ref{fig:figC12} reports HV urban driving behaviors from the CitySim dataset, including CitySim-freeC (China), CitySim-inA (non-signalized), and CitySim-inB (signalized) scenarios. Subplots \textbf{a–f} show that the shifted power law accurately represents both longitudinal and lateral acceleration distributions, reflecting stochastic interactions under complex urban traffic conditions.
Figure~\ref{fig:figC13} demonstrates the fitting results for AVs across diverse datasets, including CATS ACC, Waymo, Argoverse~2, MicroSimACC, OpenACC (ASta, ZalaZone, Casale, Vicolungo), and Central Ohio datasets. Subplots \textbf{a–j} highlight the model’s ability to generalize across platforms, geographies, and driving styles. Across all AV datasets, the fitted distributions achieve high $\mathrm{R}^2$ values (typically above 0.95), validating the model’s robustness in representing both frequent maneuvers and rare long-tail behaviors. 
Collectively, these results demonstrate that the shifted power law offers a unified and interpretable statistical representation for real-world stochasticity in both HV and AV driving, making it suitable for safety-aware simulation, risk evaluation, and policy testing at scale.

\subsection{Fitting results with fixed scale}\label{seca: a5_results}

Figures~\ref{fig:figC21}–\ref{fig:figC23} present the fitting results when the scale parameter is fixed at $a=5$ for both HV and AV datasets. This controlled experiment tests whether a shared scaling factor can still reproduce the diverse stochastic behaviors observed across highway, urban, and autonomous driving contexts. Despite constraining the scale, the shifted power law achieves consistently high consistency with empirical data across all scenarios, as reflected by $\mathrm{R}^2$ values comparable to those obtained in the free-parameter case. The fitted curves accurately capture both the central distribution and long-tail deviations in acceleration for the highD and CitySim datasets (HV) as well as for multiple AV datasets, including CATS ACC, Waymo, Argoverse~2, MicroSimACC, and OpenACC. These results confirm that the exponent parameter $k$ predominantly governs the behavioral variability, while the fixed scale $a=5$ still enables strong generalization across distinct platforms, regions, and driving conditions. Overall, the fixed-scale fitting further validates the robustness, interpretability, and transferability of the shifted power law for characterizing real-world stochastic driving behavior. It should be noted that we only consider the cases with $\mathrm{R}^2 > 0.8$ to calculate the risk index.

\newpage
\section{Measurement results}

In this section, we present more measurement details and analyze the results. Specifically, we first compare the results between the Gaussian distribution and the proposed shifted power law in \ref{seca: metrics}. In addition, in \ref{seca: long_tail}, we compare the shifted power law with a few typical long-tail distributions to further demonstrate the long-tail nature of driving behavior.

\subsection{Results of different metrics}\label{seca: metrics}

Tables \ref{taba: metrics_highD_lat} and \ref{taba: metrics_highD_lon} report the results of the highD dataset. For the lateral direction, the Gaussian distribution yields extremely large RP5 values, whereas the shifted power law consistently maintains values close to one across all locations. This suggests that the Gaussian assumption severely underestimates the tail behavior, while the shifted power law provides a more balanced fit. In terms of log-likelihood, the shifted power law also achieves substantially higher values compared to the Gaussian’s constant, indicating that the shifted power law captures the empirical distribution more effectively. Regarding KL divergence, the shifted power law achieves lower or comparable divergence at most locations, with the average value reduced from 0.344 to 0.190.

A similar trend is observed in the longitudinal direction. The Gaussian model again produces disproportionately large RP5 values, while the shifted power law remains well-bounded around 0.8. The log-likelihood improvement is consistent, showing that the shifted power law aligns better with the data. For KL divergence, the shifted power law achieves reductions at Location 2 and Location 5. Nevertheless, the averaged KL divergence across all locations is comparable between the two methods (0.312 for Gaussian versus 0.292 for shifted power law), highlighting that the proposed distribution achieves better fidelity while maintaining numerical stability.

Compared with the highD dataset, the CitySim results show a similar advantage of the shifted power law as listed in Tables \ref{taba: metrics_CitySim_lat} and \ref{taba: metrics_CitySim_lon}. In the lateral direction, the shifted power law yields substantially higher log-likelihood values and reduces the KL divergence at Intersection B and Freeway C. On average, the shifted power law decreases RP5 by several orders of magnitude and lowers the KL divergence from 0.604 to 0.362. In the longitudinal direction, the shifted power law consistently outperforms the Gaussian across all three metrics, with the average KL divergence reduced from 0.385 to 0.133.

The AV datasets further demonstrate the robustness of the shifted power law, as listed in Table~\ref{taba: metrics_AV}. Across all datasets, the Gaussian model exhibits extremely large RP5 values, whereas the shifted power law remains close to one. The log-likelihood values again highlight the better fit of the shifted power law, with improvements of more than 200 on average. KL divergence is consistently reduced, with the average decreasing from 0.494 under Gaussian to 0.118 under the shifted power law. Notably, in large-scale datasets such as Waymo and MicroSimACC, the shifted power law achieves obvious reductions in both RP5 and KL divergence, underscoring its ability to generalize across diverse autonomous driving datasets.

The comparison confirms that the shifted power law, as a representative long-tail distribution, provides a more accurate and stable prediction than the Gaussian baseline commonly adopted in prior studies. Because modeling HV and AV driving behavior fundamentally involves long-tailed dynamics, it is also important to compare the shifted power law with other common long-tail distributions to demonstrate its advantage.

\subsection{Comparison with long-tail distributions}\label{seca: long_tail}

To further validate the long-tail nature of HV and AV driving behavior, we compare the proposed shifted power law with a few typical long-tail distributions, including Laplace and Student's t distributions. The standardized PDFs of these two distributions are represented as Equations \eqref{eqa: Laplace} and \eqref{eqa: Student}, respectively.
\begin{equation}\label{eqa: Laplace}
    f^{\text{Laplace}}(x) = \frac{1}{\sqrt{2}}\exp{(-\sqrt{2}|x|)},
\end{equation}
\begin{equation}\label{eqa: Student}
    f^{\text{Student}}(x) = \frac{\Gamma(\frac{\nu + 1}{2})}{\sqrt{(\nu-2)\pi} \Gamma(\frac{\nu}{2})}\left(1+\frac{x^2}{\nu-2}\right) ^ {-(\nu+1)/2},
\end{equation}
where $\nu$ denotes the degree of freedom.

Tables~\ref{taba: metrics_others_highD}–\ref{taba: metrics_others_AV} summarize the results. Compared with Laplace and Student’s t, the shifted power law consistently achieves RP5 values close to one, indicating stable tail fitting, while the other distributions often yield much larger deviations. The shifted power law also provides higher log-likelihoods and generally lower KL divergence across datasets. For example, in the AV datasets, its average KL divergence is 0.118, substantially smaller than Laplace (0.201) and Student’s t (0.206–0.262). These findings demonstrate that even among long-tail models, the shifted power law offers the most accurate and robust characterization of HV and AV driving behavior.

\newpage
\section{Comparison of different prediction strategies}\label{seca: prediction_comparison}

To evaluate how prediction strategies influence the fidelity and efficiency of behavioral modeling under the proposed shifted power law, we first compare the distributions of the normalized residual among different prediction horizons (i.e., 0.2\,s, 0.4\,s, 0.6\,s, and 0.8\,s) in \ref{seca: prediction_horizon}. In addition, we also compare the performance between one-step prediction and multi-step prediction in \ref{seca: multi_step_prediction}. For the scope of this analysis, and to conserve computational resources, these experiments are conducted exclusively on the highD dataset.

\subsection{Comparison among different prediction horizons}\label{seca: prediction_horizon}

A key consideration in trajectory prediction is the temporal granularity of the forecast. In this study, we configure our model with a base time step of $\Delta t = 0.2~\text{s}$ over a maximum of $T=12$ steps. We also evaluate the prediction performance over different forward horizons: 0.2\,s, 0.4\,s, 0.6\,s, and 0.8\,s. To ensure a fair comparison where each prediction horizon is derived from the same quantity of past context, the number of input time steps $T$ is adjusted accordingly. Specifically, for prediction horizons of 0.2\,s, 0.4\,s, 0.6\,s, and 0.8\,s, the input sequence lengths are set to $T=12$, $6$, $4$, and $3$, respectively. This configuration ensures that all models observe a consistent 2.4-second previous horizon, isolating the effect of the forward prediction horizon.

The results are visualized in Figures~\ref{figa: com_pred_steps_lat} and \ref{figa: com_pred_steps_lon}. As anticipated, predictions made over longer forward horizons (e.g., 0.8\,s) exhibit a slightly more dispersed distribution compared to their shorter-horizon counterparts. This increase in dispersion reflects the accumulated uncertainty in forecasting dynamic systems further into the future. Nevertheless, the overall shape and trend of the distributions across the different horizons remain remarkably consistent. This indicates that the model's predictive capability scales in a stable manner with the horizon length. Consequently, a 0.2\,s single-step prediction provides a robust and representative basis for evaluating model performance.

\subsection{Comparison between one-step prediction and multi-step prediction}
\label{seca: multi_step_prediction}

In addition to the single-step analysis, we evaluate the model's performance in a multi-step prediction setting. This is a more challenging and practical task, as it requires the model to recursively forecast a sequence of future states. The multi-step prediction is formally defined as:
\begin{equation}\label{eqa: multi_step_pre}
    \hat{y}_{T+1:T+z}, \hat{\gamma}_{T+1:T+z} = g^{\mathrm{AI}}(x_{1:T})
\end{equation}
where \( z \) is the number of future time steps to be predicted. For each future time step \( i \), the normolzed residual is calculated as:
\begin{equation}
    \bar{\sigma}_{T+i} = \frac{y_{T+i} - \hat{y}_{T+i}}{\hat{\gamma}_{T+i}}, \quad \forall i \in \{1,2,\cdots, z\},
\end{equation}
where \( y_{T+i} \) is the ground-truth value, \( \hat{y}_{T+i} \) is the predicted mean, and \( \hat{\gamma}_{T+i} \) is the predicted standard deviation at that step. Single-step prediction provides a focused assessment of the model's immediate forecasting capability at a specific future point. On the contrary, multi-step prediction evaluates the model's average performance and its ability to maintain accuracy over an extended trajectory, as the normalized residuals from all \( z \) steps are aggregated into a single distribution for analysis. For a direct and meaningful comparison, we maintain a consistent time discretization of $\Delta t = 0.2$\,s for the multi-step predictions. This allows us to directly juxtapose the normalized residual distributions of single-step and multi-step predictions at identical prediction horizons—namely, 0.2\,s, 0.4\,s, 0.6\,s, and 0.8\,s. A single-step prediction of 0.8\,s is achieved in one large stride (\(i=4\)), while a multi-step prediction to the same horizon is achieved through four recursive 0.2\,s steps (\(z=4\)).

Figures~\ref{figa: com_pred_methods_2_lat}–\ref{figa: com_pred_methods_8_lon} provide pairwise comparisons between the one-step recursive prediction and the multi-step prediction at different rollout intervals (0.2–0.8\,s). 
At the shortest interval (0.2\,s; Figures~\ref{figa: com_pred_methods_2_lat} and~\ref{figa: com_pred_methods_2_lon}), the one-step formulation reproduces almost identical central and tail regions compared with the multi-step prediction across all six locations, demonstrating strong equivalence in stochastic representation.
As the temporal interval increases to 0.4\,s and 0.6\,s (Figures~\ref{figa: com_pred_methods_4_lat}–\ref{figa: com_pred_methods_6_lon}), the overall distributional patterns remain well aligned, with only slight broadening of the tails in denser flow conditions (e.g., Locations~4–6). 
Even at the longest prediction interval (0.8\,s; Figures~\ref{figa: com_pred_methods_8_lat} and~\ref{figa: com_pred_methods_8_lon}), the one-step approach continues to capture both the central density and the heavy-tail decay of the multi-step prediction, confirming its robustness against temporal drift and compounding error.

Overall, the expanded results across Locations~1–6 confirm that the one-step recursive formulation achieves nearly identical statistical outcomes to the multi-step strategy in both longitudinal and lateral dimensions. The consistency across all time horizons and sites supports the use of the one-step prediction for long-term behavioral simulation, offering superior computational efficiency and real-time compatibility while maintaining high fidelity under the shifted power law framework.

\newpage

\section{Simulation implementation}\label{seca: simulation}

To empirically evaluate the performance of the trained behavior prediction models, we implement a high-fidelity agent-based simulation environment. Each simulation rollout is initialized by sampling seed frames from the highD dataset, from which the initial states for all vehicles are constructed. The state vector $x_{1:T}$ for each vehicle and its neighbors encompasses key kinematic and contextual variables, including positions, velocities, and lane assignments. At every time step, the trained behavior prediction model $g^{\mathrm{AI}}$ provides next-step mean value and standard deviation $(\hat{y}_{T+1}, \hat{\gamma}_{t+1})$. The full acceleration distribution is then constructed as $\tilde{y}_{T+1} = \hat{y}_{T+1} + \hat{\gamma}_{t+1} \cdot \bar{\sigma}$, where $\bar{\sigma}$ follows a standard Gaussian distribution $\mathcal{N}(0,1)$ for a baseline Gaussian model or the proposed shifted power law as defined in Equation~\eqref{eq:scaling_pdf}.

The physical acceleration for each vehicle is sampled from this reconstructed distribution:
\begin{equation}
    a_{T+1} \sim \tilde{y}_{T+1}.
\end{equation}
This sampled acceleration is used to propagate the vehicle's state forward in time. The velocity $v_{T+1}$ and location $x_{T+1}$ are updated according to the following kinematic equations:
\begin{equation}
    v_{T+1} = v_{T} + a_T \cdot \Delta t + \frac{1}{2}(a_{T+1} - a_T)\cdot \Delta t,
\end{equation}
\begin{equation}
    x_{T+1} = x_T + v_{T}\cdot \Delta t + \frac{1}{2}a_{T} \cdot (\Delta t)^2 + \frac{1}{6}(a_{T+1} - a_T) \cdot (\Delta t)^2,
\end{equation}
where $\Delta t$ is the time duration between two time steps (i.e., 0.2\,s). The state of the simulation environment, including the updated surrounding vehicle information for each agent, is subsequently refreshed to prepare for the next time step. At the conclusion of each simulation run, key performance metrics are recorded, including the Vehicle Miles Traveled (VMT) and the number of collisions for each vehicle.

The simulation experiments are executed on a workstation equipped with an AMD Ryzen Threadripper PRO 7985WX CPU, 2$\times$ NVIDIA RTX 6000 Ada GPUs, and 256 GB memory, running Ubuntu 22.04 LTS. Each simulation batch consists of multiple parallel rollouts to cover a total driving distance of approximately 10 million km, which requires approximately four days of continuous computation under this configuration.

\end{document}